\documentclass[aps,prd,superscriptaddress,onecolumn,floatfix,nofootinbib,preprintnumbers]{revtex4}

\usepackage{amsmath,graphicx,tabularx,url,color}

\begin{document}
\def\tablename{Table}
\def\figurename{Figure}

\setlength{\floatsep}{0pt}
\setcounter{topnumber}{1}
\setcounter{bottomnumber}{1}
\setcounter{totalnumber}{2}

\def\gtot{\Gamma_\text{tot}}
\def\brinv{\text{BR}_\text{inv}}
\def\brsm{\text{BR}_\text{SM}}
\def\bratio{\mathcal{B}_\text{inv}}
\def\as{\alpha_s}
\def\az{\alpha_0}
\def\gz{g_0}
\def\w{\vec{w}}
\def\sdag{\Sigma^{\dag}}
\def\s{\Sigma}
\newcommand{\psib}{\overline{\psi}}
\newcommand{\Psib}{\overline{\Psi}}
\newcommand\one{\leavevmode\hbox{\small1\normalsize\kern-.33em1}}
\newcommand{\Mpl}{M_\mathrm{Pl}}
\newcommand{\p}{\partial}
\newcommand{\lag}{\mathcal{L}}
\newcommand{\qqquad}{\qquad \qquad}
\newcommand{\qqqquad}{\qquad \qquad \qquad}

\newcommand{\qb}{\bar{q}}
\newcommand{\matx}{|\mathcal{M}|^2}
\newcommand{\really}{\stackrel{!}{=}}
\newcommand{\msbar}{\overline{\text{MS}}}
\newcommand{\qns}{f_q^\text{NS}}
\newcommand{\lqcd}{\Lambda_\text{QCD}}
\newcommand{\met}{\slashchar{p}_T}
\newcommand{\ptmiss}{\slashchar{\vec{p}}_T}
\newcommand{\pmiss}{\slashchar{p}}

\newcommand{\st}[1]{\tilde{t}_{#1}}
\newcommand{\stb}[1]{\tilde{t}_{#1}^*}
\newcommand{\nz}[1]{\tilde{\chi}_{#1}^0}
\newcommand{\cp}[1]{\tilde{\chi}_{#1}^+}
\newcommand{\cm}[1]{\tilde{\chi}_{#1}^-}

\providecommand{\mg}{m_{\tilde{g}}}
\providecommand{\mst}{m_{\tilde{t}}}
\newcommand{\msn}[1]{m_{\tilde{\nu}_{#1}}}
\newcommand{\mch}[1]{m_{\tilde{\chi}^+_{#1}}}
\newcommand{\mne}[1]{m_{\tilde{\chi}^0_{#1}}}
\newcommand{\msb}[1]{m_{\tilde{b}_{#1}}}

\newcommand{\mev}{{\ensuremath\rm MeV}}
\newcommand{\gev}{{\ensuremath\rm GeV}}
\newcommand{\tev}{{\ensuremath\rm TeV}}
\newcommand{\fb}{{\ensuremath\rm fb}}
\newcommand{\ab}{{\ensuremath\rm ab}}
\newcommand{\pb}{{\ensuremath\rm pb}}
\newcommand{\sign}{{\ensuremath\rm sign}}
\newcommand{\ifb}{{\ensuremath\rm fb^{-1}}}

\def\slashchar#1{\setbox0=\hbox{$#1$}           
   \dimen0=\wd0                                 
   \setbox1=\hbox{/} \dimen1=\wd1               
   \ifdim\dimen0>\dimen1                        
      \rlap{\hbox to \dimen0{\hfil/\hfil}}      
      #1                                        
   \else                                        
      \rlap{\hbox to \dimen1{\hfil$#1$\hfil}}   
      /                                         
   \fi}
\newcommand{\dslash}{\slashchar{\partial}}
\newcommand{\Dslash}{\slashchar{D}}

\def\eg{{\sl e.g.} \,}
\def\ie{{\sl i.e.} \,}
\def\etal{{\sl et al} \,}

\title{Tagging single Tops}

\author{Felix Kling}
\affiliation{Institut f\"ur Theoretische Physik, Universit\"at Heidelberg, Germany}

\author{Tilman Plehn}
\affiliation{Institut f\"ur Theoretische Physik, Universit\"at Heidelberg, Germany}

\author{Michihisa Takeuchi}
\affiliation{Institut f\"ur Theoretische Physik, Universit\"at Heidelberg, Germany}

\begin{abstract}
 Top taggers which identify and reconstruct boosted top quarks have
 been established as novel tools for a multitude of LHC analyses. We
 show how single top production in association with a light-flavor or
 bottom jet can be observed in the boosted phase space regime. The full top
 reconstruction as part of the tagging algorithm allows us to define a
 distinctive kinematic angle which clearly separates different single
 top production processes.
\end{abstract}

\maketitle

\tableofcontents

\newpage

\section{Introduction}
\label{sec:intro}

Because of its large mass the top quark offers a unique handle on the
structure of electroweak symmetry breaking and possible links to the
origin of flavor. Its properties, like mass, charge, or $W$-helicity
fractions are mainly measured in top pair production with subsequent
top decays~\cite{topmass,topcharge,whelicity}. The charged-current
$tbW$ coupling is directly accessible in single top
production, \ie without correlation to the remaining CKM mixing
structure.  The combined Tevatron analysis sets a lower limit on it as
$V_{tb} >0.77$ with 95\% C.L., consistent with
$V_{tb}=1$~\cite{Group:2009qk}.

At the LHC we can study three different single top production modes:
$s$-channel $tb$ production via a time-like virtual $W$-boson,
$t$-channel $tq$ production via a space-like virtual $W$-boson, and
$tW$ production in association with a real $W$-boson.  All of them
should and will be separately measured, to test the electroweak
properties of the heavy third quark generation.  New physics
contributions to the $tbW$ coupling or to any of these three processes
could be a first step to discover physics beyond the Standard
Model~\cite{bsm_review,Tait:2000sh,Alwall:2006bx}. General FCNC couplings, heavy
$W^\prime$ gauge bosons, or a forth
generation are only a few
examples.\medskip

In this context, it is worth to note that CDF reports a 2.5~$\sigma$
deviation from the Standard Model prediction in the ratio between
$s$-channel and $t$-channel cross sections; only the sum of both channels
agrees with the Standard Model predictions~\cite{singletop_CDF}. The
corresponding D0 results are consistent with the Standard
Model~\cite{singletop_D0}.  Recently, both experiments updated their
results and are still inconsistent with each other at the 3~$\sigma$
level~\cite{CDFsingletop75,D0singletop54}. An improved understanding
of single top production at the LHC seems at order.\medskip

The main difference between single top production at the Tevatron and
at the LHC is that the $s$-channel production rate is significantly
smaller than $t$-channel rate at higher collider energies. This is
because of the large gluon content in the proton which mainly enhances
$t$-channel production (through $g\to b\bar{b}$ splitting) and $tW$
production.

\begin{table}[b]
\hfill
\begin{tabular}{l|rrrrrr}
\hline
8 TeV: $p_{T,t}^\text{min}$& 0 & 100 & 200 & 300 & 400 & 500 \cr
\hline
$\sigma_s$~[fb] &5548 &1784 &349 & 86.4 & 26.5 & 9.54 \cr
$\sigma_t$~[fb] & 86829& 18167 & 2273 & 409.2 & 95.7   &  26.0 \cr
$\sigma_{t\bar{t}}$~[fb] & 234731 & 137274 & 34640 &  7560 &1850 &  519 \cr
\hline
$\sigma_s/\sigma_t (\%)$ & 6.4 & 9.8 & 15.4 & 21.1 & 27.7 & 36.7 \cr
$\sigma_s/\sigma_{t\bar{t}} (\%)$  &2.36 & 1.29& 1.00 & 1.14 & 1.43 & 1.83  \cr
\hline
\end{tabular}
\hfill
\begin{tabular}{l|rrrrrr}
\hline
14 TeV: $p_{T,t}^\text{min}$   & 0 & 100 & 200 & 300 & 400 & 500 \cr
\hline
$\sigma_s$~[fb]       &11852 &4206 &964 & 292 & 108 & 43.8 \cr
$\sigma_t$~[fb]      & 248194& 59621 & 9128 & 2038 & 583   &  203 \cr
$\sigma_{t\bar{t}}$~[fb]      & 917935 & 572517 & 167564 &  43700 &12771 &4304 \cr
\hline
$\sigma_s/\sigma_t (\%)$     & 4.7 & 7.0 & 10.5& 14.3 & 18.5 & 21.5 \cr
$\sigma_s/\sigma_{t\bar{t}}(\%) $  & 1.23  & 0.73 & 0.57 & 0.66 & 0.85 & 1.07  \cr
\hline
\end{tabular}
\hfill
\caption{Top production cross sections for different minimum $p_{T,t}$
  values for 8~TeV (left) and 14~TeV (right) center-of-mass energy.
  For $t\bar{t}$ we require at least one top exceeding the minimum
  $p_{T,t}$.}
\label{tab:crosssection}
\end{table}

The relative size of the two different production rates strongly
depends on the transverse momentum of the top quark.  In
Tab.~\ref{tab:crosssection} we show the different cross sections for
top production with a variable minimum transverse momentum of the top
quark.  We gain a factor two for $s$-channel production relative to
$t$-channel production when we focus on events with
$p_{T,t}>200$~GeV.\medskip

\begin{figure}[t]
\includegraphics[width=0.32\textwidth]{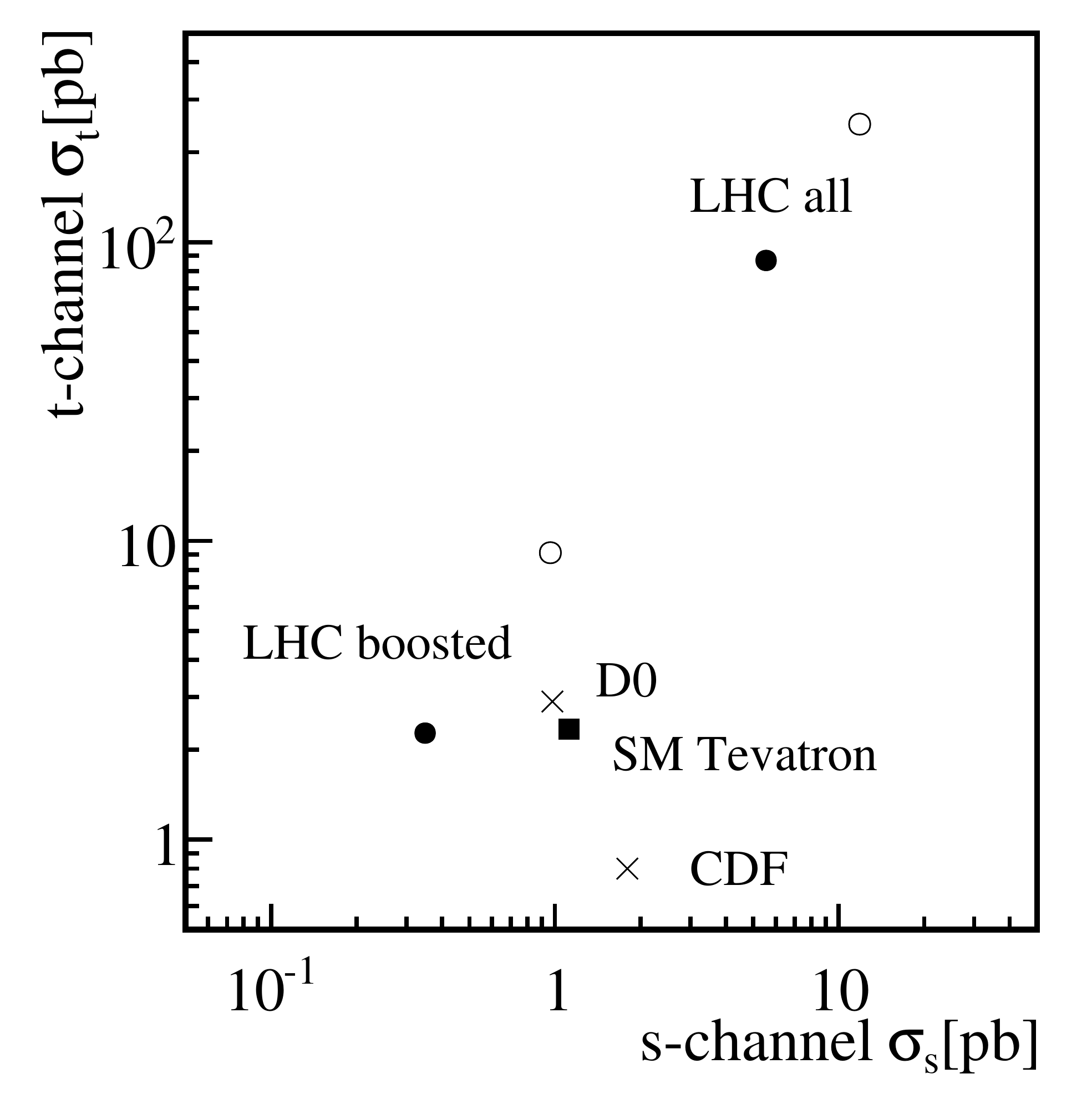}
\hspace{0.1\textwidth}
\includegraphics[width=0.32\textwidth]{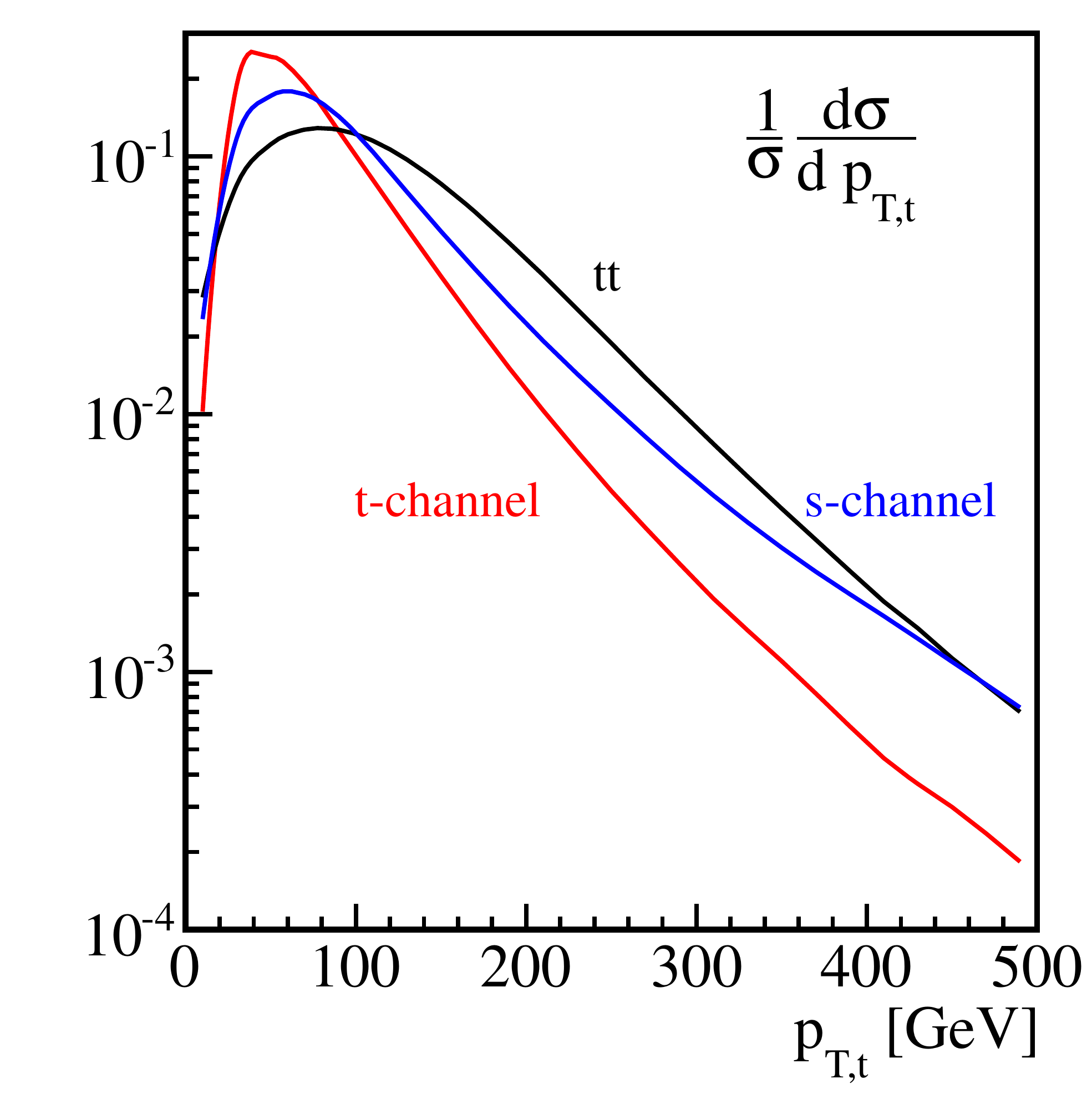}
\caption{Left: $\sigma_s$ vs. $\sigma_t$ at the LHC. We show cross
  sections for inclusive single tops at the LHC, boosted tops
  ($p_{T,t} > 200~\gev$) at the LHC, and Tevatron. The open (filled)
  circles correspond to 14~TeV (8~TeV) results.  Right: normalized
  $p_{T,t}$ distributions for $s$-channel, $t$-channel single top and
  $t\bar{t}$ at the 14~TeV LHC.}
\label{fig:st}
\end{figure}

Measuring cross sections only in this boosted $p_{T,t}$ range provides
independent information about single top production, in addition to the
fully inclusive measurement.  The left panel of Fig.~\ref{fig:st}
shows both single top cross sections $\sigma_s$ vs. $\sigma_t$ for all
tops and for boosted tops only at 14~TeV (open circles) and at 8~TeV
(filled circles).  The CDF and D0 measurements together with the
Standard Model prediction for the Tevatron are included for reference.
The ratio of $s$-channel to $t$-channel cross sections becomes larger
for boosted tops.  The right panel of Fig.~\ref{fig:st} shows the
corresponding $p_{T,t}$ distributions for 14~TeV.  Indeed, the
$s$-channel curve drops slower than the other production channels.
The only down side of the boosted regime is that the $t\bar{t}$
background is enhanced.  This happens because $t\bar{t}$ production is
mainly gluon-initiated and spin conservation suppresses the threshold
regime.\medskip

In recent years, top tagging algorithms using jet substructure have
rapidly
matured~\cite{toptagger1,toptagger2,toptagger3,toptagger4,toptagger5,toptagger6,
 boost_proceedings,tagger_review}.  A particularly efficient top
tagging algorithms for moderately boosted hadronic tops is the {\sc
  HEPTopTagger}~\cite{HEP1,HEP2,HEP3,HEPstop2012,HEP_ATLAS}.  In this
paper we investigate its possible benefit for single top searches in
the fully hadronic decay mode.  We will show that tagging boosted
single tops allows us to overcome all backgrounds and extract both,
$s$-channel and $t$-channel single top production.

\section{Event generation}
\label{sec:event}

Single top production accompanied with quarks can be categorized into
\begin{alignat}{5}
pp  \to t \bar{b} \quad (s\text{-channel}) 
\qqquad \text{and} \qqquad
pp \to  t q \quad (t\text{-channel}) \; ,
\label{eq:signal}
\end{alignat}
plus the hermitian conjugate final states.
For $t$-channel production the possible final state quarks can be 
$q = d,s,\bar{u},\bar{c}$. We
treat $tW$ production as a background.  To leading order, the two
channels in Eq.\eqref{eq:signal} are obviously well separated.
Overlapping contributions of the kind $qg \to t\bar{b} q^\prime$ do
appear at next-to-leading order (NLO) contributions to both
processes. These are $s$-channel diagrams where one of the initial
quarks is provided by gluon splitting, or $t$-channel diagrams where
one of the initial $b$-quarks arises through gluon splitting.  They do
not interfere because of different color flows; the $t\bar{b}$ system
forms a color singlet for the $s$-channel process and a color octet
for the $t$-channel.  Hence, to next-to-leading order the two
processes are well defined.  Diagrams with additional gluons start
interfering at NNLO level.  We consider this numerically irrelevant
complication beyond the scope of our paper, even though in full QCD
the separation of Eq.\eqref{eq:signal} should be reviewed.\medskip

As Standard Model backgrounds we consider $t\bar{t}$+jets, QCD jets,
$W$+jets, and $tW$ production~\cite{ATLASsingletop, CMSsingletop}.
All corresponding samples are generated with {\sc
  Alpgen+Pythia}~\cite{alpgen,pythia}. Supersymmetric backgrounds we
neglect for the time being. For the signal with the very hard cuts
(only) on the leading two constituents, leading order simulations with
parton shower are sufficient. For all background processes other than
$tW$ we use MLM matching~\cite{mlm} to account for hard jet
radiations.  This includes up to $t\bar{t}$+2~jets, $W$+4~jets and
$3-5$ QCD jets.  For $tW$ production we combine $tW$ and $tWb$ samples
and explicitly veto the phase space region $|m_{Wb} - m_t|<5$~GeV
overlapping with $t\bar{t}$ production (as recommended by {\sc
  Alpgen}). Eventually, we find that $tW$ production is significantly
suppressed compared to the $t\bar{t}$ background, so its simulation
details do not affect our analysis.

All single top samples we then normalize to the approximate NNLO rates
of 87.2~pb~($t$-channel), 5.55~pb~($s$-channel) and
22.2~pb~($tW$-channel) at 8~TeV~\cite{kidonakis}.  Single top and
anti-top production in the $s$-channel contribute $t:3.79$~pb and
$\bar{t}:1.76$~pb. For the $t$-channel we find $t:56.4$~pb and
$\bar{t}:30.7$~pb, and for the $tW$-channel there is no preference for
either charge.  At 14~TeV LHC the rates become 11.86~pb~($t:7.87$~pb
and $\bar{t}:3.99$~pb) for the $s$-channel, 248~pb~($t:154$~pb and
$\bar{t}:94$~pb) for the $t$-channel and 83.6~pb for the $tW$-channel.
The dominant $t\bar{t}$+jets background sample we normalize to the
approximate NNLO rate of 234~pb (918~pb) for 8~TeV
(14~TeV)~\cite{tt_nnlo}.  For the remaining sub-leading background
samples we use the leading order normalization.\medskip

Our detailed analysis includes {\sc Delphes} with default ATLAS
detector setting as a fast detector simulation~\cite{delphes}.  The
calorimeter cell information provided by {\sc Delphes} is used as
(fat)-jet constituents. As usual, we rely on the Cambridge/Aachen
(C/A) algorithm~\cite{ca_algo} implemented in {\sc
  FastJet}~\cite{fastjet}.  The resulting fat jets are then used as
input for the {\sc HEPTopTagger}~\cite{HEP1,HEP3}.  The same C/A
algorithm we use for regular QCD jets, but with $R=0.5$. For regular
well separated jets the algorithm should not matter, though.  All
leptons we require to be hard and isolated: $p_{T,\ell} > 10$ GeV and
no track of another charged particle within $R < 0.5$ around the
lepton, based on {\sc Delphes}. Triggering the hadronic single top
events with highly energetic fat jets might or might not be a
challenge, which we unfortunately have to leave to a more detailed
experimental analysis.

\section{Single tops at 8~TeV}
\label{sec:8tev}

In this section we discuss all selection cuts which we apply in our
single top analyses and show results for $t$-channel and $s$-channel
single top production at 8~TeV.  We classify the set of cuts into
three classes:
\begin{enumerate}
\item cuts on the tagged top jet
\item cuts on the balanced top and recoil jet system
\item cuts on the recoil jet
\end{enumerate}

\subsection{Top tag}
\label{sec:toptag}

The starting point of our analysis is a balanced system of a fat top
jet and its high-momentum recoil system.  Hence, vetoing isolated
leptons we first require two fat jets
\begin{equation}
p_{T,\text{fat}}>200~\gev \; . 
\label{eqcut:fatjet}
\end{equation}
In those two fat jets we require exactly one top tag to avoid
$t\bar{t}$ background.  We use the {\sc HEPTopTagger} algorithm with
modified parameters setting: the top mass window we slightly narrow to
$[160,200]$~GeV instead of the default $[150, 200]$~GeV. Similarly, we
reduce the $W$ mass window to $\pm 10\%$ instead of $\pm 15\%$, and
increase the lower soft-collinear mass cut to $\arctan(m_{13}/m_{12})>
0.45$ instead of 0.2.  The tighter cuts reduce the top tagging
efficiency but increase the fake-top rejection. Relative to the
required two fat jets we now find a tagging efficiency of $12 - 13$\%
for signal events, and $16\%$ for $t\bar{t}$ events. The latter is
higher because there are two hadronic tops in each event but less than
twice the single top efficiency because more events have two fat jets.
The fake top rate for QCD sample is about 1\%, \ie 0.5\% per fat jet.
The fake rate for $W$+jets is about 3\%.  Both fake rates are based on
the samples after requiring at least 3 jets with $R=0.4$ in an
event.\medskip

\begin{figure}[b]
\includegraphics[width=0.32\textwidth]{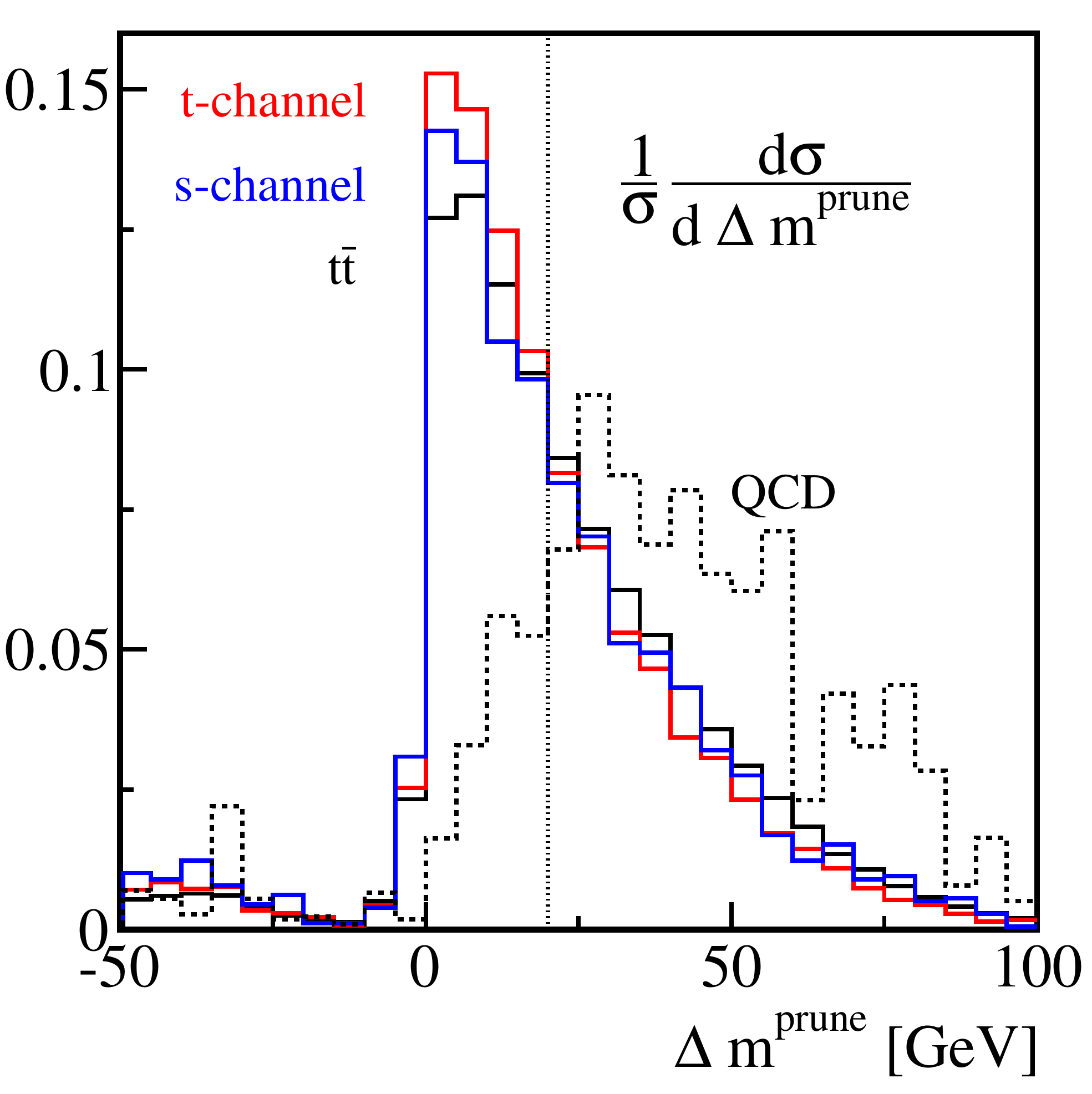}
\caption{Pruned-filtered mass difference $\Delta m^\text{prune}$ distributions from the
  top tagger for the signal and the leading backgrounds.}
\label{fig:prune}
\end{figure}

To extract the single top signal from the overwhelming QCD background we
can use the pruned mass~\cite{pruning,HEP3} in addition to the
filtered mass~\cite{bdrs}.  Figure~\ref{fig:prune} shows their
differences $\Delta m^\text{prune} = m^\text{prune} -
m^\text{filter}$.  We impose a cut
\begin{equation}
-10 < \Delta m^\text{prune} <20~\gev \; , 
\label{eqcut:prune}
\end{equation}
which is passed by about half of the events including tops and about
1/6 left for QCD jets. For the same purpose we then require a $b$-tag
inside the top tag.  A $b$-tagging efficiency of 50\%~\cite{giacinto}
translates into specific $40\%$ inside the top tag, since the correct
$b$-subjet selection ranges around 80\%~\cite{HEP3}.  For QCD and
$W$+jets we assume a 1\% fake rate.\medskip

In lines 0-4 of Tab.~\ref{tab:t8tev} we show the corresponding cut
flow for signal and backgrounds. The set of cuts on the tagged top
alone has an efficiency of $2 - 3$\% for samples including tops and
$1.5\times 10^{-5}$ for QCD relative to the number of events with two fat jets.  
Top pair production and QCD jets are the
two main backgrounds at this stage.

\begin{table}[t]
\centering
\begin{tabular}{l|rr|rrrr||rr}
\hline
8 TeV: rates in fb & $t$-channel  & $s$-channel & $t\overline{t}$ & $tW$ & QCD & $W+$jets& $S/B$ & $S/\sqrt{B}_{10 \ifb}$\\
\hline
0. cross section   &8.72$\cdot 10^4$ &5.55$\cdot 10^3$ &2.34$\cdot 10^5$ &4.06$\cdot 10^4$&6.58$\cdot 10^8$&1.57$\cdot 10^6$ &-- &--\\ 
1. $n_\ell=0$ with 2 fat jets  [Eq.\eqref{eqcut:fatjet}] &1.57$\cdot 10^3$ &230 &1.88$\cdot 10^4$ &1.63$\cdot 10^3$  &6.67$\cdot 10^6$&4.81$\cdot 10^4$  &0.0002 &1.9 \\ 
\hline
2. one top tag    &204 &28.2 &3070 &227  &6.38$\cdot 10^4$ &1297  &0.003 &2.5 \\  
3. $\Delta m^\text{prune}$ cut  [Eq.\eqref{eqcut:prune}] 
&110 &13.9 &1421 &102  & 9.71$\cdot 10^3$ &530  &0.009 &3.2\\
4. $b$-tag in top tag  
& 44.3 &     5.29 &      524 &     37.4 &     97.1 &      5.30 &   0.07 & 5.4\\
\hline
5. $p_{tj}$ cut [Eq.\eqref{eqcut:tj}] 
& 15.3 &     1.34 &     11.1 &     1.12 &     12.4 &     1.27 &    0.57 & 9.3\\
6. $\cos \theta^*<-0.5$   [Eq.(\ref{eqcut:cost})]     
& 8.6 &   0.07 &     1.58 &    0.14 &      3.3 &    0.21 &     1.62 & 11.8\\
\hline
\end{tabular}
\caption{Cut flow for the single top analysis at 8~TeV. The
  significances are quoted for $t$-channel single top production
  assuming all other processes as backgrounds.}
\label{tab:t8tev}
\end{table}

\subsection{Top-jet system: t-channel}

Once the top is tagged we turn to the recoil jet which provides enough
boost to the top. We ignore all calorimeter cells used for the
constituents of the tagged top and cluster the remaining entries using
the C/A jet algorithm with $R=0.5$. We select the hardest jet as the
(leading) recoil jet and require it to be above $p_{T,j}> 25$~GeV and
inside the second fat jet.  This way we define a reconstructed top
momentum and its recoil.\medskip

\begin{figure}[b]
\includegraphics[width=0.32\textwidth]{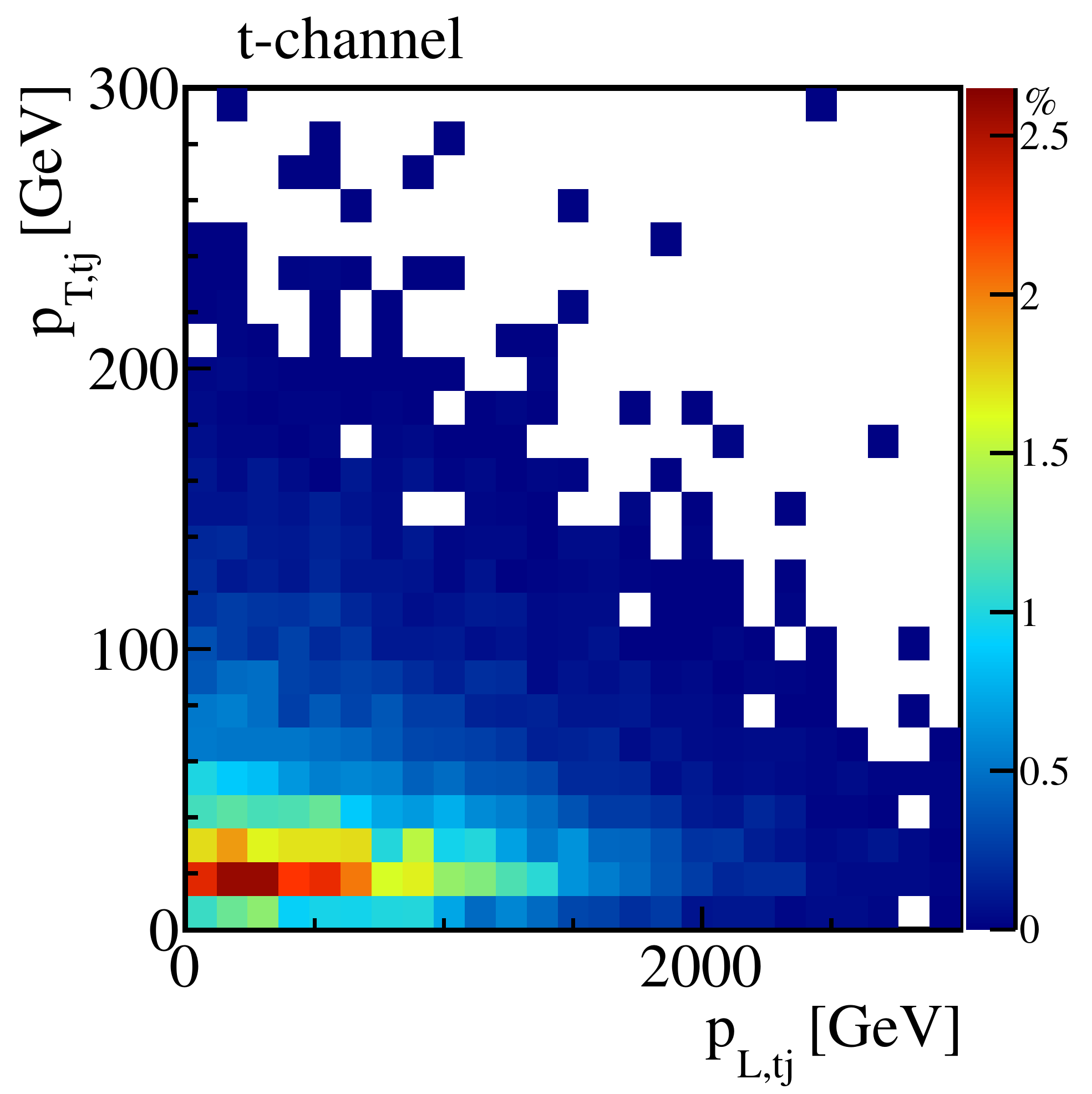}
\includegraphics[width=0.32\textwidth]{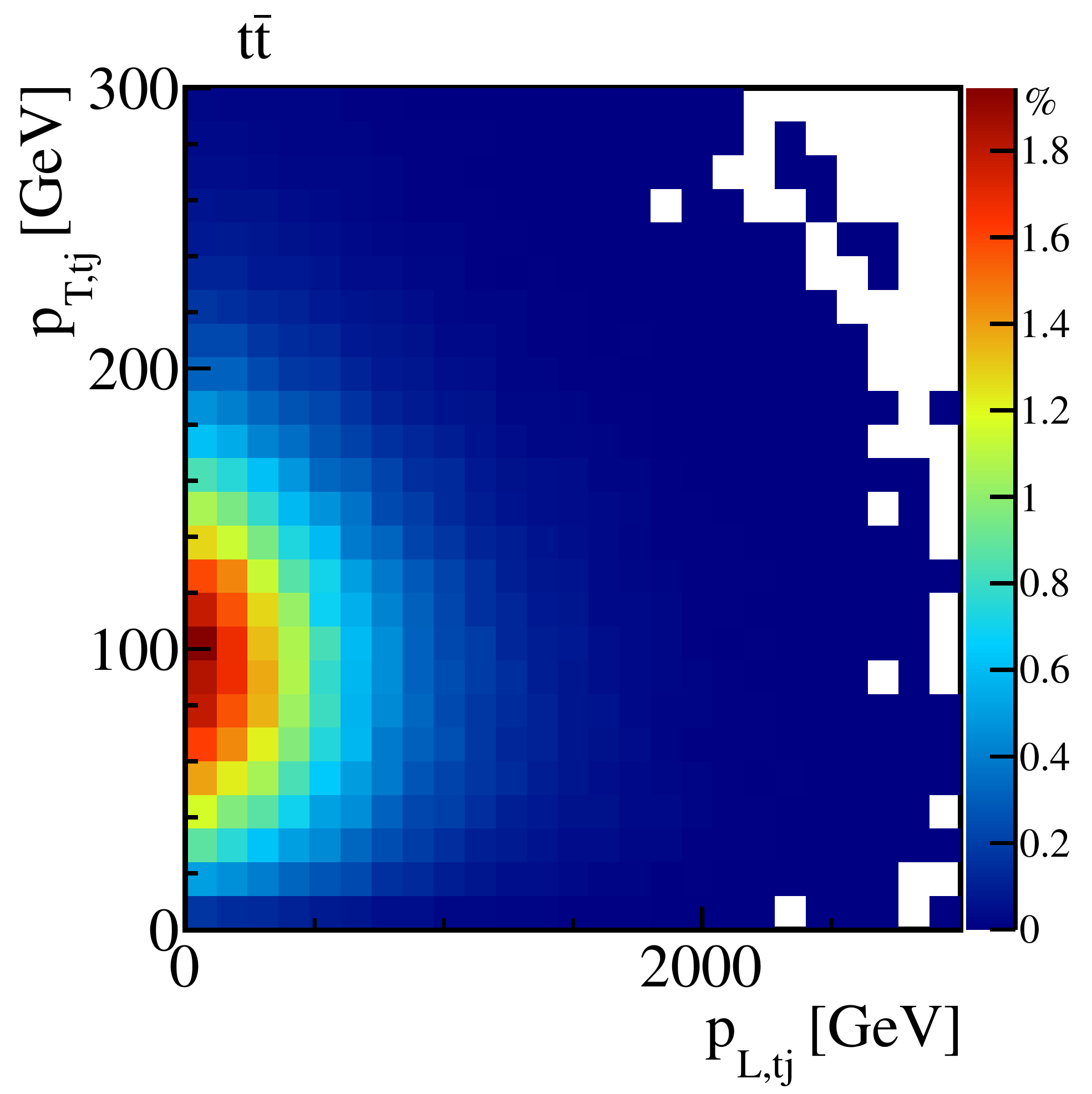}
\includegraphics[width=0.32\textwidth]{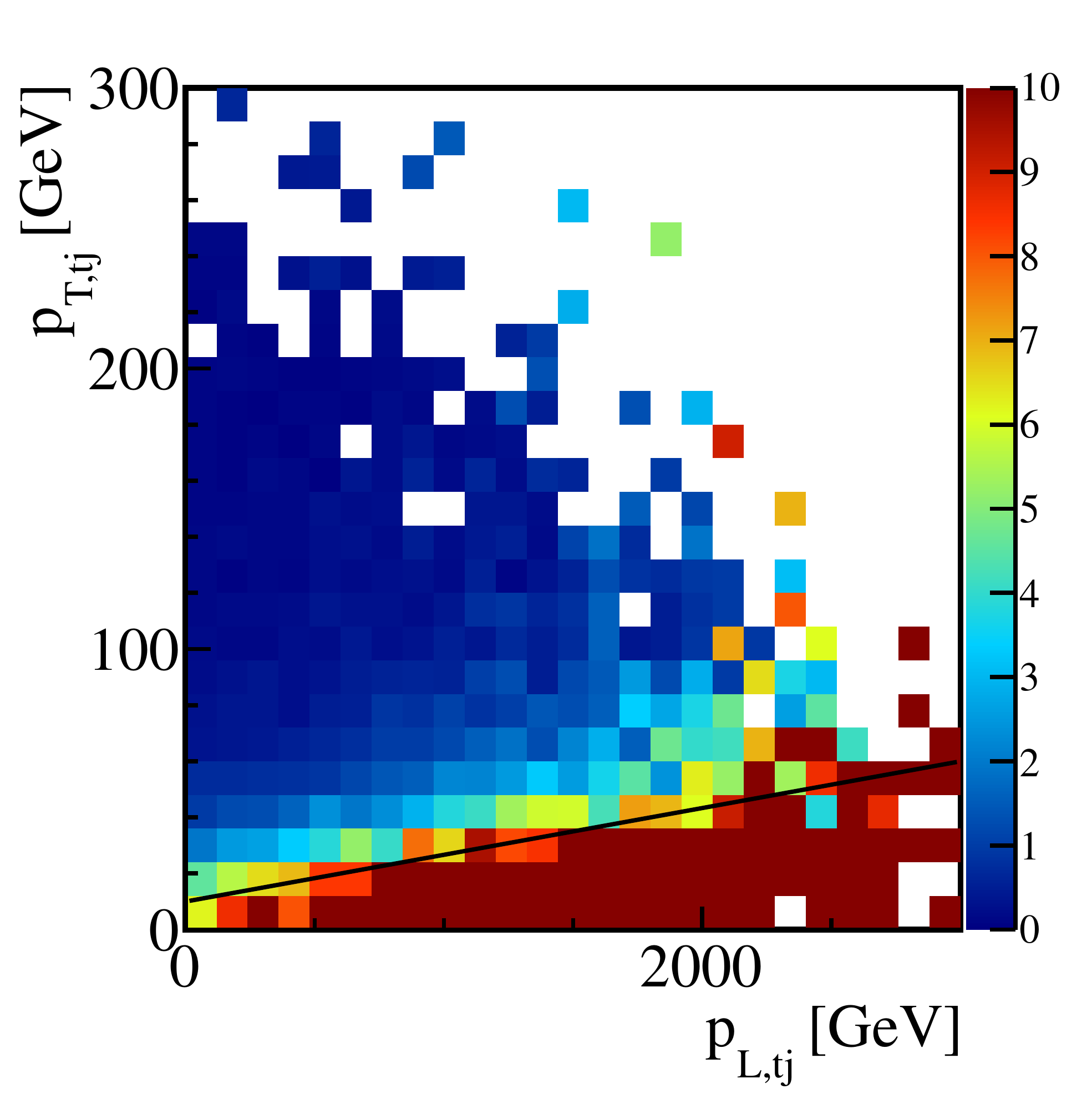}
\caption{$p_{L,tj}$ vs. $p_{T,tj}$ distributions for $t$-channel
  single top production, $t\bar{t}$ production, and their ratio at 8~TeV.}
\label{fig:system}
\end{figure}

The first variable we look at is the top-jet system momentum.  The
goal is to reject the leading $t\bar{t}$ background at this stage.
Figure~\ref{fig:system} shows the longitudinal vs. the transverse system
momentum for $t$-channel single top and $t\bar{t}$ production.  We
observe a distinct difference in their ratio: for the signal
$p_{T,tj}$ tends to be small and $p_{L,tj}$ large while the opposite
is true for top pairs.  This can be understood for $t\bar{t}$ remembering that two
pair-produced tops start off back-to-back, but the selected recoil jet
often only includes part of the second top. The longitudinal system
momentum is generally small for the dominant gluon fusion
process. For the signal it can be understood by the unbalanced valence quark
initial states.

Based on Fig.~\ref{fig:system} we enhance the single top samples
relative to top pairs by requiring
\begin{equation}
p_{T,tj} < \frac{p_{L,tj}}{60} + 10~\gev \; .
\label{eqcut:tj}
\end{equation}
This cut rejects up to 98\% of the $t\bar{t}$ and $tW$ backgrounds
while keeping 35\% of the signal. It is less effective against QCD
(87\%) but still helpful.  This is because QCD events are dominated by
di-jets, one of which fakes the tagged top.  The $s$-channel single
top events behave similar to the $t$-channel and survive to
$25\%$. All results are shown in Tab.\ref{tab:t8tev}.

As a side remark, the transverse component of the system momentum is
similar to the Collins-Soper angle
$\arctan(p_{T,tj}/m_{tj})$~\cite{collins_soper}.  Smaller $p_{T,tj}$
corresponds to a smaller Collins-Soper angle, but in our case the
correlation between $p_{L,tj}$ and $p_{T,tj}$ appears to be the more
powerful cut.\medskip

Initial-state parton combinations contribute differently to the
different processes.  The participating $b(\bar{b})$-parton density is
even softer than for the other sea quarks; the system moves into the
valence quark direction.  Initial state radiation affects this
argument to some degree, but the main feature should be visible.  Our
second variable reflects this topology of single top Feynman diagrams
and related kinematic enhancements.  We define $\theta^*$ as the angle
between (anti-)top momentum in the rest frame of the $tj$-system and
the boost vector $\vec{\beta}$ from the rest frame to the laboratory
frame. Many details on this angle are presented around
Fig.~\ref{fig:kinematics} in the appendix.

\begin{figure}[t]
\includegraphics[width=0.32\textwidth]{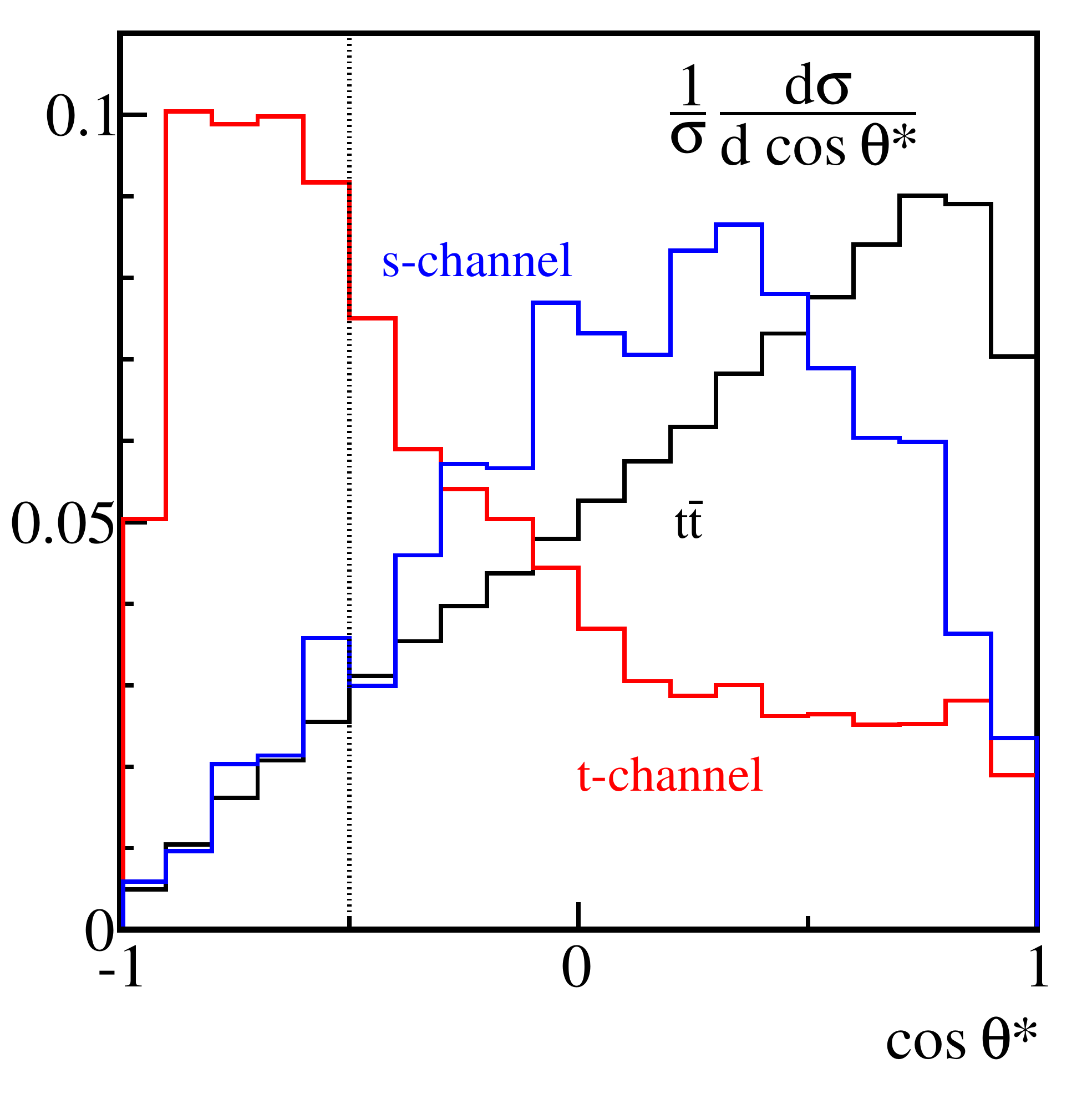}
\hspace*{0.1\textwidth}
\includegraphics[width=0.32\textwidth]{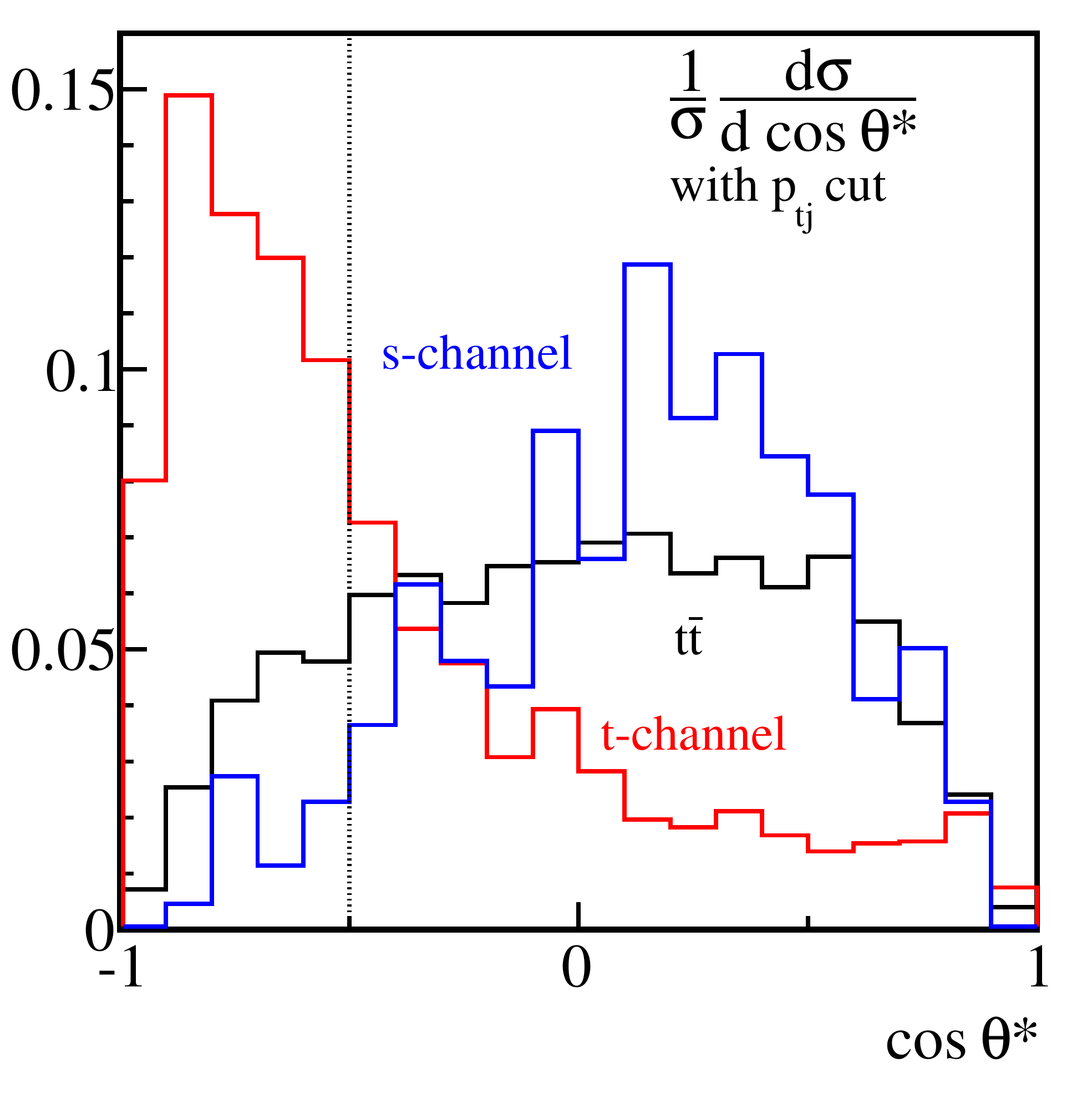}
\caption{Normalized $\cos \theta^*$ distributions before~(left) and
  after~(right) the $p_{tj}$ cut for $t$-channel single top,
  $s$-channel single top, and top pair production at 8~TeV.}
\label{fig:cos}
\end{figure}

Figure~\ref{fig:cos} shows the $\cos \theta^*$ distributions after all
cuts to step~4 and to step~5 but without the cut on $\Delta m^\text{prune}$.
We checked that this cut does not affect the shown distributions.
The $\cos \theta^*$ distributions for
single top production are reflecting the (polar) scattering angle
distributions. This is because the direction of $\vec{\beta}$ follows
the direction of the harder initial-state partons.
For $t$-channel single top production  with a $q b$ initial state
the top tends to be emitted in the direction of the $b$-quark, 
corresponding to $\cos\theta^* \sim -1$.
The same preference we expect from $t$-channel anti-top production.

For $s$-channel single top production with a $u\bar{d}$ initial state
the top tends to be emitted in the direction of the valence $u$-quark,
\ie around $\cos \theta^* \sim 1$. The lower peak position can be
understood from the decreasing top tagging efficiency once the top
starts overlapping with the beam or has low $p_{T,t}$.  For $s$-channel anti-top 
production, $d\bar{u}$ is the main initial parton combination and it essentially
results in the opposite $\cos \theta^*$ distribution.  Provided
that we cannot distinguish top charge in hadronic mode, the
distribution we observe is combined distribution and contaminated by
smaller anti-top distribution.

The singly tagged top pair system moves toward the top direction,
because usually the $tj$-system only includes part of the second top
as the assumed recoil. We indeed see a clear preference of large
values $\cos \theta^* \sim 1$.  Because this is the same reason as we
already quoted for the transverse momentum balance of $tj$-system this
feature vanishes once we apply the cut Eq.\eqref{eqcut:tj}. In the
right panel of Fig.~\ref{fig:cos} the top pair distribution is
essentially flat.  Finally, for QCD there exists no clear correlation
from the di-jet topology.\medskip

We can turn this argument into the single top selection cut
\begin{equation}
\cos \theta^* < -0.5 \qquad (t\text{-channel}) \; .
\label{eqcut:cost}
\end{equation}
As shown in Tab.\ref{tab:t8tev} this cut leaves us with $S/B=1.62$ and
$S/\sqrt{B}>10$ for 10 $\ifb$. One might expect a $b$-veto in the
recoil as part of the $t$-channel single top search strategy, but
since at this stage we are dominated by the pure QCD background we
refrain from it.

\subsection{Top recoil structure: s-channel} 

For the $s$-channel single tops we apply the same event selection as
shown up to step~5 in Tab.~\ref{tab:t8tev}.  The only difference is
that the recoil jet is actually a central yet un-tagged bottom jet.
In Fig.~\ref{fig:cos} we see that for $s$-channel single top
production we should require
\begin{equation}
\cos \theta^* > -0.5 \qquad (s\text{-channel}) \; .
\label{eqcut:coss}
\end{equation}
The numbers after successive cuts for $s$-channel search are
summarized in Tab.~\ref{tab:s8tev}.  The dominant backgrounds then
are $t$-channel single top, $t\bar{t}$, and QCD jets.\medskip

\begin{table}[t]
\centering
\begin{tabular}{l|rr|rrrr||rr}
\hline
8~TeV: rates in fb & $t$-channel  & $s$-channel & $t\overline{t}$ & $tW$ & QCD & $W+$jets& $S/B$ & $S/\sqrt{B}_{25 \ifb}$\\
\hline
1-5. one top tag, $b$-tag, $p_{tj}$ cut  [Eq.\eqref{eqcut:tj}] 
& 15.3 &     1.34 &     11.1 &     1.12 &     12.4 &     1.27 &    --  & -- \\
6. $\cos \theta^*>-0.5$     [Eq.\eqref{eqcut:coss}]      &
6.75 &     1.27 &     9.52 &    0.97 &     9.06 &     1.06 &   0.05 & 1.2 \\
\hline
7. $b$-tag in recoil jet   &
0.07 &    0.64 &    1.94 &   0.18 &  0.09 &  0.01 &   0.28 & 2.1\\
8. $E_j^{R<0.2}/E_\text{fat}$, \ $m_j < 65$~GeV [Eq.\eqref{eqcut:ejratio02}]  & 
0.04 &   0.35 &   0.11 &  0.02 &  0.03 & -- &    1.75 & 3.9 \\
9. $\pmiss_T < 40$~GeV [Eq.\eqref{eqcut:pmiss}]  &
0.04 &   0.32 &  0.07 &  0.02 &  0.03 & --  &    2.00 & 4.0  \\
\hline
\end{tabular}
\caption{Cut flow for the $s$-channel single top analysis at 8~TeV.
  The significance values are quoted for $s$-channel single top
  production assuming all other channels as backgrounds and 
  an integrated luminosity of 25~$\ifb$.}
\label{tab:s8tev}
\end{table}

The first additional cut on the recoil jet should obviously be a
$b$-tag.  Because for top pairs the probability of identifying the
$b$-jet with the leading recoil jet is far from 100\% this cut is
efficient also against the $t\bar{t}$ background.  We again assume a
50\% $b$-tagging efficiency and 1\% fake rate.  After this
requirement, the $t$-channel single top, QCD, and $W$+jets backgrounds
are under control.

\begin{figure}[b]
\includegraphics[width=0.32\textwidth]{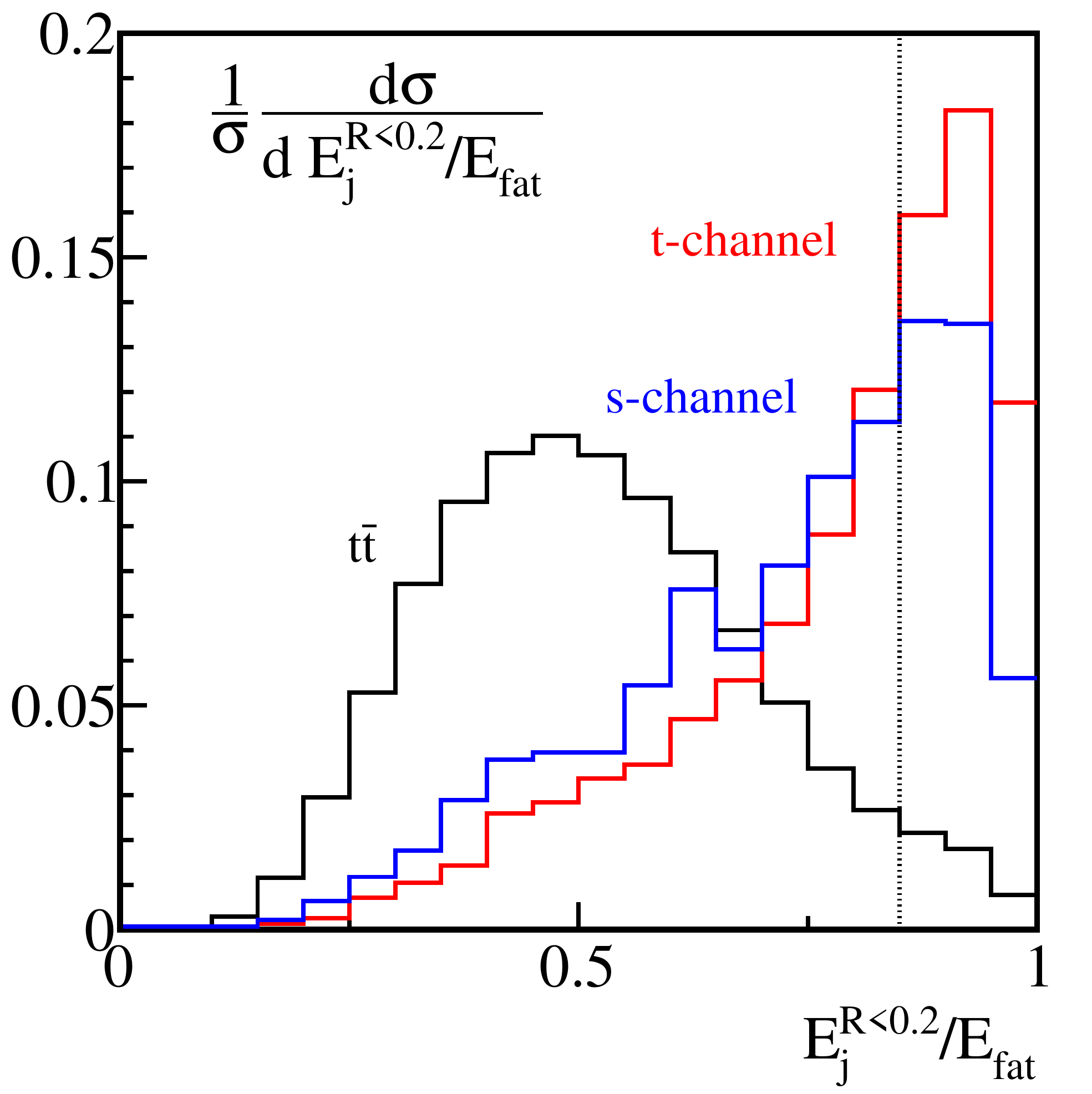}
\caption{ $E_j^{R<0.2}/E_\text{fat}$ distributions for each process,
  as defined before Eq.\eqref{eqcut:ejratio02}.}
\label{fig:ejef}
\end{figure}

To better reject the $t\bar{t}$ background we use the fact that the
recoiling fat jet in single top production should be narrower and more
isolated.  For $t\bar{t}$ pairs the fat jet corresponding to the
recoil jet often contains more than one sub-jet. To quantifying this
feature, we define the energy fraction of the filtered (leading)
recoil jet inside the fat jet $E_j^{R<0.2}/E_\text{fat}$, where
$E_j^{R<0.2}$ is the filtered energy of the recoil jet (with
$R_\text{filter}=0.2$ and $n_\text{filter}=1$) and $E_\text{fat}$ is
the energy of the fat jet which contains the recoil jet.  This
condition is very similar to the usual lepton isolation criterion.
Figure~\ref{fig:ejef} shows the $E_j^{R<0.2}/E_\text{fat}$ distributions.  Top
pairs indeed lead to a softer distribution, not peaked at unity.
Hence, we require
\begin{equation}
\frac{E_j^{R<0.2}}{E_\text{fat}} > 0.85 
\qqquad \text{and} \qqquad  m_j < 65~\gev \; ,
\label{eqcut:ejratio02}
\end{equation}
where $m_j$ is the jet mass of the recoil jet. Recoil jets consistent
with a boosted $W$-decay are also removed by the second condition
$m_j < 65~\gev$, even though the fraction is not large.  This cut reduces
$t\bar{t}$ events to roughly $5\%$ while keeping half of the signal. We
also considered a cut on $E_j/E_\text{fat}$ without filtering instead,
but it gives a weaker background suppression.\medskip

At this stage, a part of $t\bar{t}$ background consists of events
with a leptonic ($e, \mu, \tau$) top decay. This means it includes
large missing momentum which we can use to require
\begin{equation}
\pmiss_T < 40~\gev \; .
\label{eqcut:pmiss}
\end{equation}
Combining all set of the cuts $s$-channel signal analysis for 8~TeV
LHC results in $S/B=2.0$ and $S/\sqrt{B}=4.0$ for 25~$\ifb$.

\section{Single tops at 14 TeV} 
\label{sec:14tev}

Following our results for the 8~TeV LHC run we study the same two
signatures for 14~TeV collider energy. Because our analysis relies on
boosted and hence fairly energetic tops we expect significant
improvements from this energy increase.  All relevant cross sections
are of course larger. Following Tab.~\ref{tab:crosssection} the
boosted top cross sections increase by roughly a factor 3 for
$s$-channel single top production, 4 for the $t$-channel, and 5 for
$t\bar{t}$.  The latter implies that there will be a tradeoff between
$S/B$ and the significance from an improved boosted regime.\medskip

\begin{table}[b]
\centering
\begin{tabular}{l|rr|rrrr||rr}
\hline
14 TeV: rates in fb & $t$-channel  & $s$-channel & $t\overline{t}$ & $tW$ & QCD & $W+$jets& $S/B$ & $S/\sqrt{B}_{10 \ifb}$\\
\hline
0. cross section 
&2.48$\cdot 10^5$ &1.18$\cdot 10^4$ &9.20$\cdot 10^5$ &1.60$\cdot 10^5$ &1.94$\cdot 10^9$&3.88$\cdot 10^6$  &0.0003 & -- \\ 
1. $n_\ell=0$ with 2 fat jets [Eq.\eqref{eqcut:fatjet}]
&6590 &670 &9.53$\cdot 10^4$ &1.02$\cdot 10^4$  & 2.83$\cdot 10^7$ &1.29$\cdot 10^5$  &0.0004  & -- \\ 
\hline
2. one top tag        
&819   &81.4   & 1.48$\cdot 10^4$  &1350    &3.00$\cdot 10^5$ &3015   &0.003 &4.6\\  
3. $\Delta m^\text{prune}$ cut [Eq.\eqref{eqcut:prune}]     &
416 &     40.4 & 6438  &      578 & 3.61$\cdot 10^4$&    1005 &  0.009 & 6.3\\
4. $b$-tag       &
166 &     15.5 & 2346 &      212 &      361 &       10.1 &   0.06 & 9.7\\
\hline
5. $p_{tj}$ cut [Eq.\eqref{eqcut:tj}]     &
 67.8 &     4.28 &     72.7 &      9.20 &     75.5 &     2.53 &    0.41 & 16.7\\
6. $\cos \theta^*<-0.5$  [Eq.(\ref{eqcut:cost})]  &
 41.2 &    0.30 &     14.6 &     1.18 &     7.15 &    0.55 &     1.74 & 26.8\\
\hline
7. $E_j^{R<0.2}/E_\text{fat}$, $m_j < 65$~GeV [Eq.\eqref{eqcut:ejratio02}] &
 36.1 &    0.25 &     7.33 &    0.50 &     3.58 &    0.50 &     2.97 & 32.7\\
\hline
\end{tabular}
\caption{Cut flow for the $t$-channel single top analysis at
  14~TeV. The significance is computed based on the statistical error
  for $10~\ifb$.  }
\label{tab:t14tev}
\end{table}

First, we present the results for $t$-channel single top production.
Table~\ref{tab:t14tev} shows the rates after all cuts described in the
previous section.  Asking for exactly one top tag for events with two
fat jets we find an efficiency around 12\% for the single top samples,
15\% for top pairs, and around $1\%$ for QCD.  These values are almost
the same as for 8~TeV.  Next, we cut on the top tag, namely the pruned
mass given in Eq.\eqref{eqcut:prune} and a $b$-tag. We find an
efficiency around 20\% for the signal, 16\% for top pairs, 0.12\%
for QCD, and 0.3\% for $W$+jets.  Again, there are no big changes from
the 8~TeV analysis.

After applying the $tj$-system momentum cut of Eq.\eqref{eqcut:tj},
the signal rate is of similar order as the leading backgrounds.
Selecting events with $\cos\theta^* < -0.5$, we can extract
$t$-channel single top production with $S/B=1.7$.  Finally, we cut on
the recoil jet system, Eq.\eqref{eqcut:ejratio02}, and arrive at
$S/B\sim 3$ and a promising signal significance, indeed.\medskip

For the $s$-channel single top search we need to check that the
enhanced $t\bar{t}$ background does not pose a major problem at
14~TeV.  Following Tab.~\ref{tab:crosssection} the naive
signal-to-background estimate otherwise drops by almost a factor two.
The efficiencies of the successive 8~TeV cuts we show in
Tab.~\ref{tab:s8tev}.  After selecting $\cos\theta^* > -0.5$ and
requiring a $b$-tagged leading recoil jet we are left with 2~fb of
signal rate with a six times larger $t\bar{t}$ background.

After applying all selection cuts, 0.95 fb $s$-channel single top
signal left with the same amount of background mainly from $t\bar{t}$
and QCD.  As expected, the signal cross section at this stage is three
times the 8~TeV result while the $t\bar{t}$ background is six times
the value quoted in Tab.~\ref{tab:s8tev}.  We achieve $S/B=1.13$ and
$S/\sqrt{B}=5.2$ for $25~\ifb$. The signal-to-background ratio can be
improved at the expense of the significance simply by tightening the
different cuts.

\begin{table}[t]
\centering
\begin{tabular}{l|rr|rrrr||rr}
\hline
14 TeV: rates in fb & $t$-channel  & $s$-channel & $t\overline{t}$ & $tW$ & QCD & $W+$jets& $S/B$ & $S/\sqrt{B}_{25 \ifb}$\\
\hline
1-5. one top tag, $b$-tag, $p_{tj}$ cut  [Eq.\eqref{eqcut:tj}]        &
 67.8 &     4.28 &     72.7 &      9.20 &     75.5 &     2.53 &    -- & -- \\
6. $\cos \theta^* > -0.5 $   [Eq.(\ref{eqcut:coss})]        &
26.6 &     3.99 &     58.2 &     8.02 &     68.3 &     1.99 &   0.02 & 1.6 \\
\hline
7. $b$-tag in recoil jet     &
0.27 & 1.99 & 12.6 & 0.76 & 0.68 & 0.02 & 0.14 & 2.6 \\
8.  $E_j^{R<0.2}/E_\text{fat}$, $m_j < 65$~GeV [Eq.\eqref{eqcut:ejratio02}] &
0.15 &        1.00 &    0.75 &   0.08 &     0.26 &  -- &    0.80 & 4.5 \\
9. $\pmiss_T <40$~GeV  [Eq.\eqref{eqcut:pmiss}]   &
0.14 &    0.95  &    0.41 &   0.03 &     0.26  &  -- &    1.13 & 5.2 \\
\hline
\end{tabular}
\caption{Cut flow for the $s$-channel single top analysis at
  14~TeV. The significance assumes $25~\ifb$ with all other channels
  being backgrounds.}
\label{tab:s14tev}
\end{table}

\section{Conclusions}
\label{sec:conclusion}

Using the {\sc HEPTopTagger} single top production in the purely
hadronic channel can be observed at the LHC.  Controlling the
$t\bar{t}$ and QCD background is the key to these single top searches.
By applying successive cuts on the tagged top jet, the fully
reconstructed top and recoil system, we
achieve $S/B >1$ with $S/\sqrt{B}>10$ for the $t$-channel process at
8~TeV with $10~\ifb$.  For this result we need to gain a factor 10
against top pairs and a factor 4000 against QCD jets 
after selecting events with two fat jets.  Most of this is
provided by the top tag.  Additional cuts on the recoil jet, including
a $b$-tag and a cut on the size of the recoil jet, can extract the
$s$-channel with 4~sigma and $S/B > 1$ at the same energy 
with an integrated luminosity of $25~\ifb$.
To distinguish the two single top production modes and to
reject backgrounds we introduce a new angular observable $\theta^*$
which is highly efficient once we reconstruct the top momentum.

For the 14~TeV we can use the same analysis.  The signal-to-background
ratios are similar to the 8~TeV case for the $t$-channel and slightly
worse than the 8~TeV result for the $s$-channel.  Thanks to the larger
cross sections the significance exceeds $5\sigma$ even for the
$s$-channel process with an integrated luminosity of
$25~\ifb$.\medskip

\begin{center}
{\bf Acknowledgments}
\end{center}

We would like to thank Michael Spannowsky and all colleagues who have
continuously helped to develop and improve the {\sc HEPTopTagger}. The
Heidelberg ATLAS group and Dirk Zerwas we would like to thank for
their help with detector effects and their simulation.

\newpage
\appendix

\section{Detector effects}

In this appendix we summarize the effects of the fast detector
simulation {\sc Delphes} on the performance of the {\sc HEPTopTagger}.
In the left two panels of Fig.~\ref{fig:delphesmass} we show the top mass
and the $W$ mass distributions reconstructed in the $t$-channel top
sample.  Both are slightly smeared out by detector effects, just as
expected.  The top mass and $W$ mass peak positions do not shift
significantly. As long as the mass ranges required by the top tagger
are sufficiently large the tagging efficiency should not change.  In
particular, the mass ranges assumed in Section~\ref{sec:toptag} should
be conservative.\medskip

The right panel of Fig.~\ref{fig:delphesmass} shows the difference
between the pruned mass~\cite{pruning} and the filtered
mass~\cite{bdrs} ($\Delta m^\text{prune} =m^\text{prune}
-m^\text{filter}$) with and without detector effects.  This
distribution is smeared out significantly and shifted toward larger
values.  Hence, this additional observable introduced in
Ref.\cite{HEP3} requires an experimental study and validation.\medskip

\begin{figure}[t]
\includegraphics[width=0.32\textwidth]{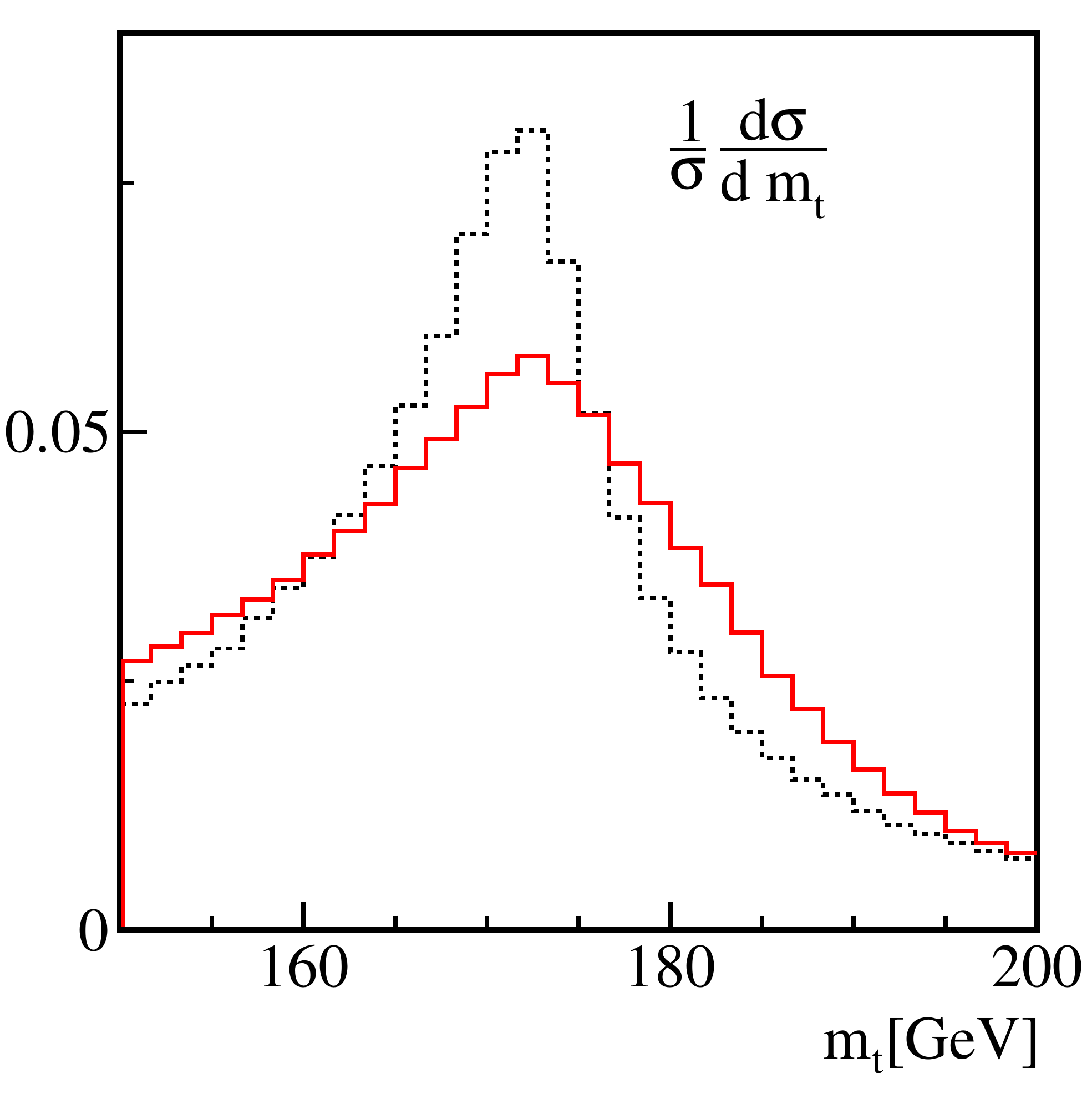}
\includegraphics[width=0.32\textwidth]{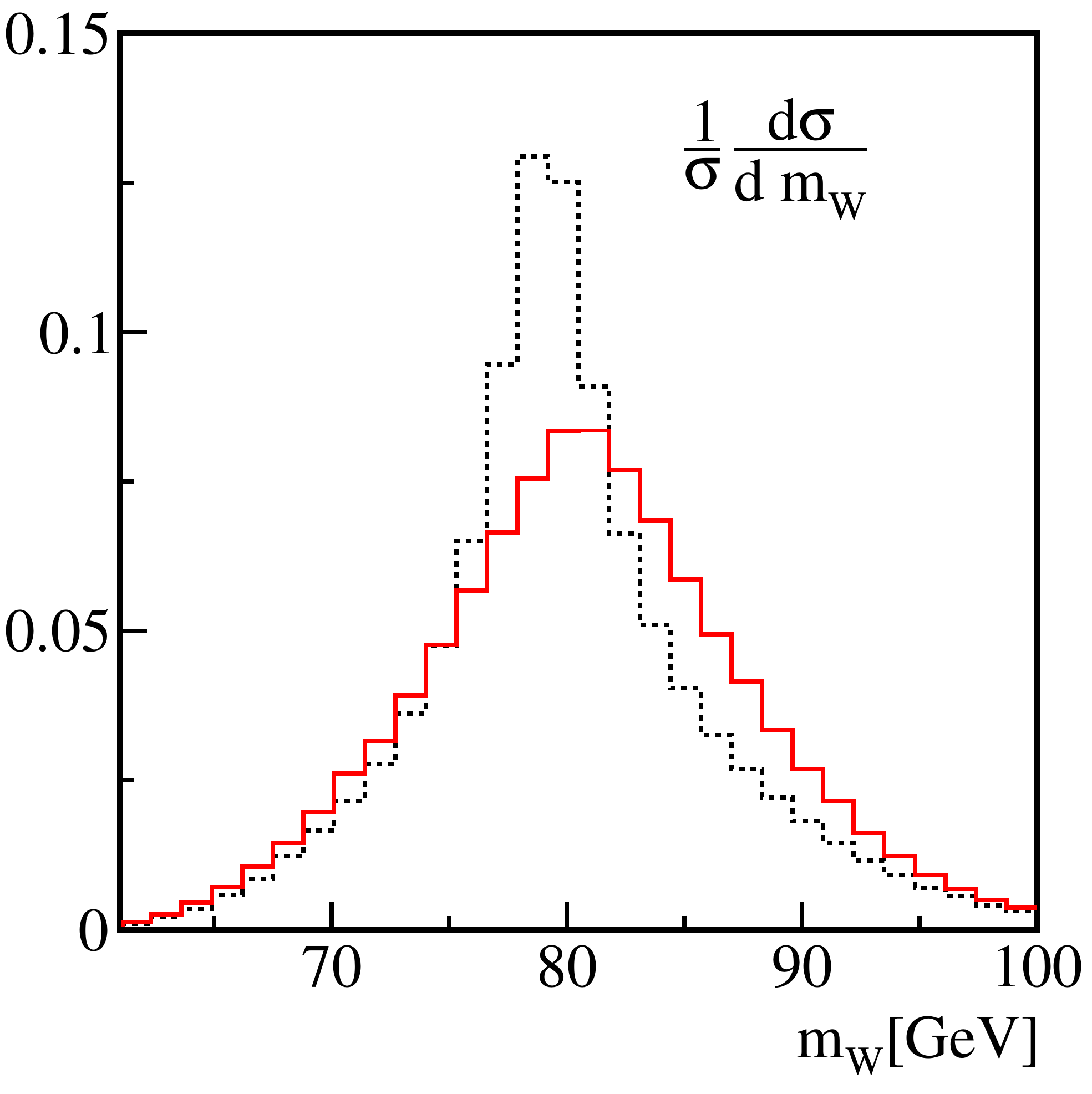}
\includegraphics[width=0.32\textwidth]{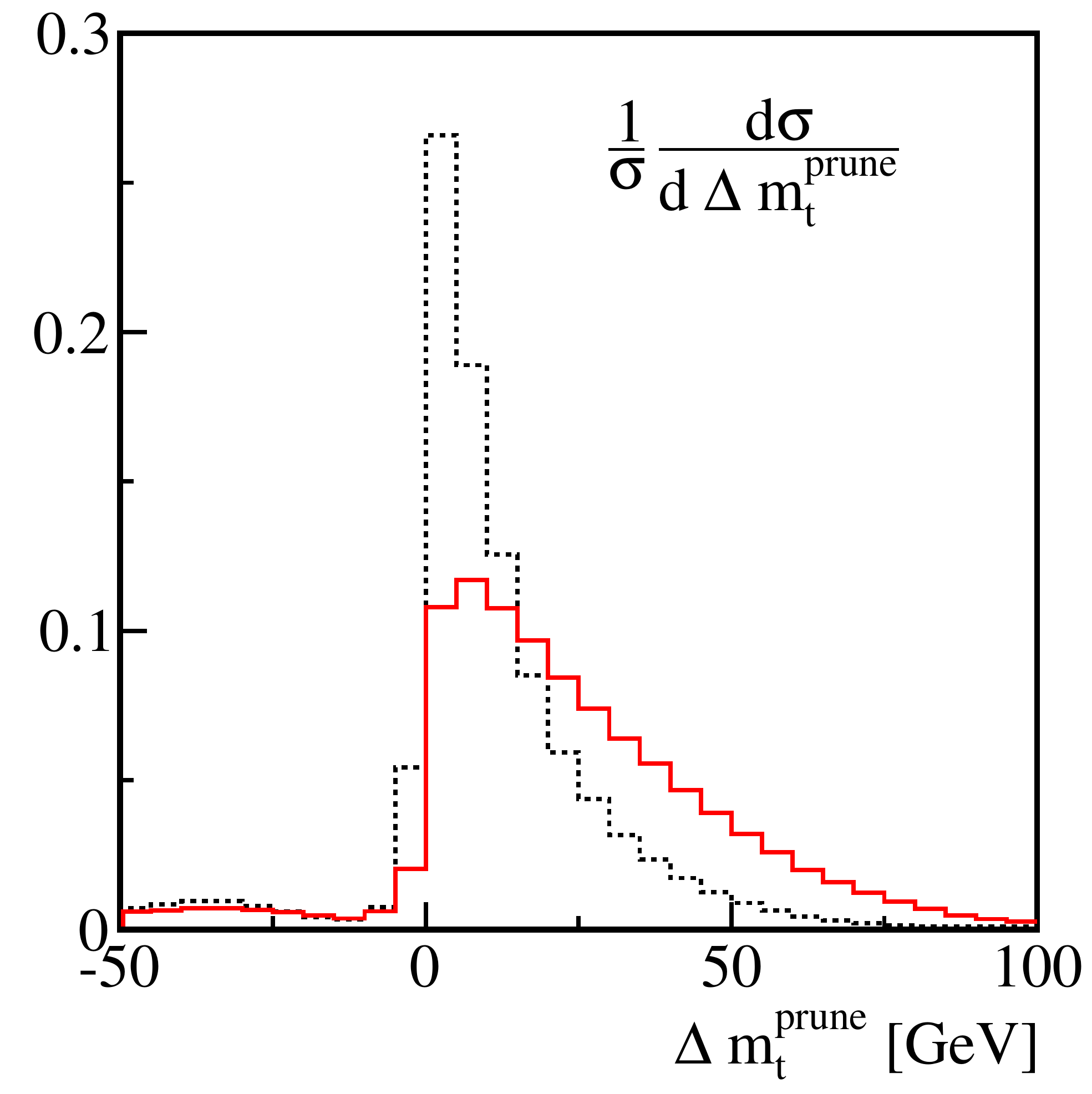}
\caption{Top tagging observables $m_t$, $m_W$, and $m_t^\text{prune}$
  for the $t$-channel single top sample with (solid red) and without
  {\sc Delphes} (dashed black).}
\label{fig:delphesmass}
\end{figure}

Figure~\ref{fig:delphes} shows how well the top tagger reconstructs
the top momentum. Detailed particle-level results including the
quality of the actual tagging algorithm can be found in
Ref.~\cite{HEP3}. In this appendix we focus on its {\sc Delphes}
simulation.  We show, from left to right, 
$\Delta p_t = |\mathbf{p}_t^\text{tag} -\mathbf{p}_t|$, 
$\Delta p_{T,t} = p_{T,t} ^\text{tag}- p_{T,t}$,
$\Delta p_t/|p_t^\text{tag}|$,
$\Delta p_{T,t}/p_{T,t}^\text{tag}$, 
and $\Delta R$.  We see slight smearing but no significant qualitative
difference.  Most tagged tops are reconstructed within an error bar of
$\Delta R <0.2$ and within a 15\% error in the momentum $p_t$.

\begin{figure}[b]
\includegraphics[width=0.195\textwidth]{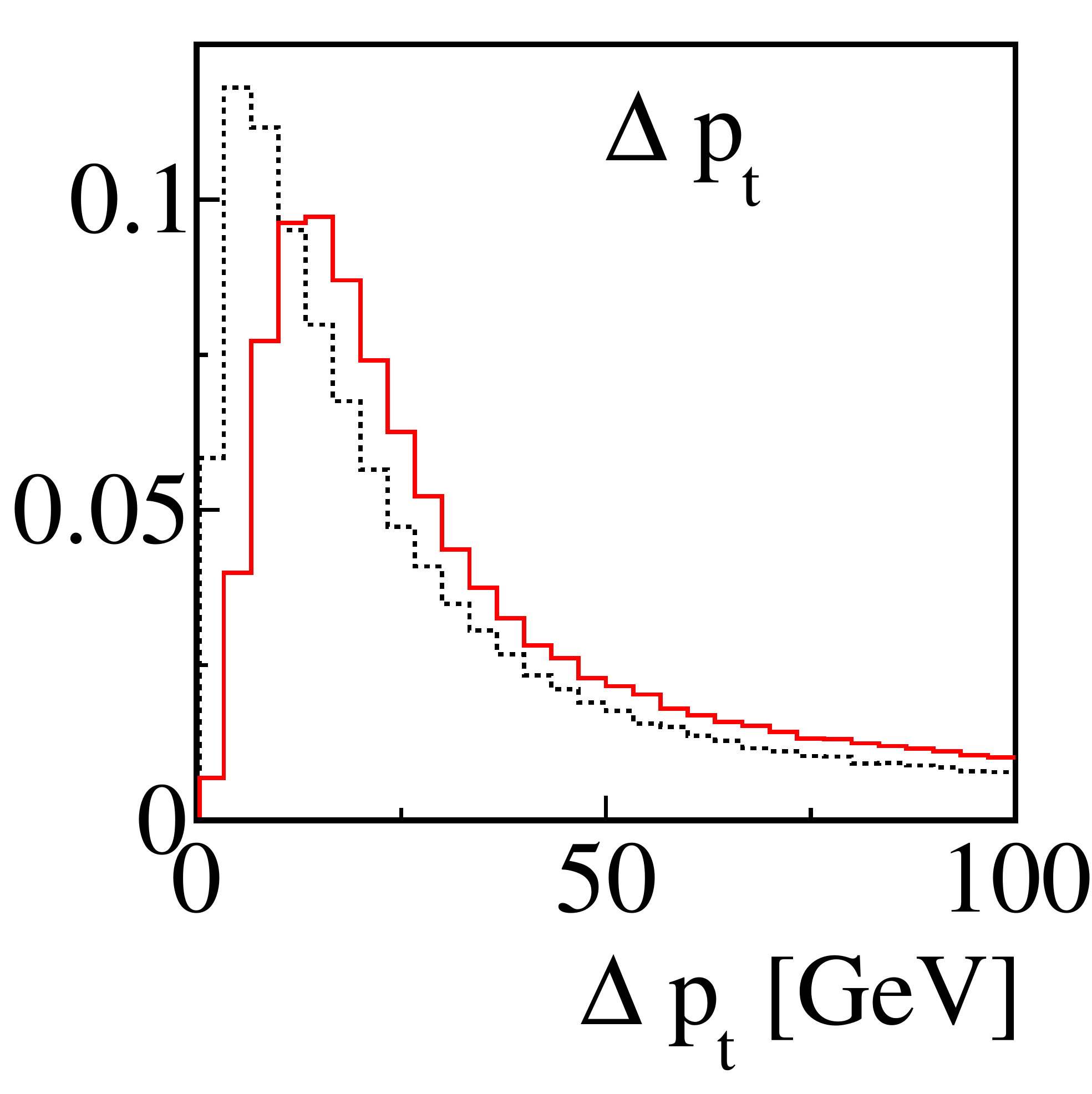}
\includegraphics[width=0.195\textwidth]{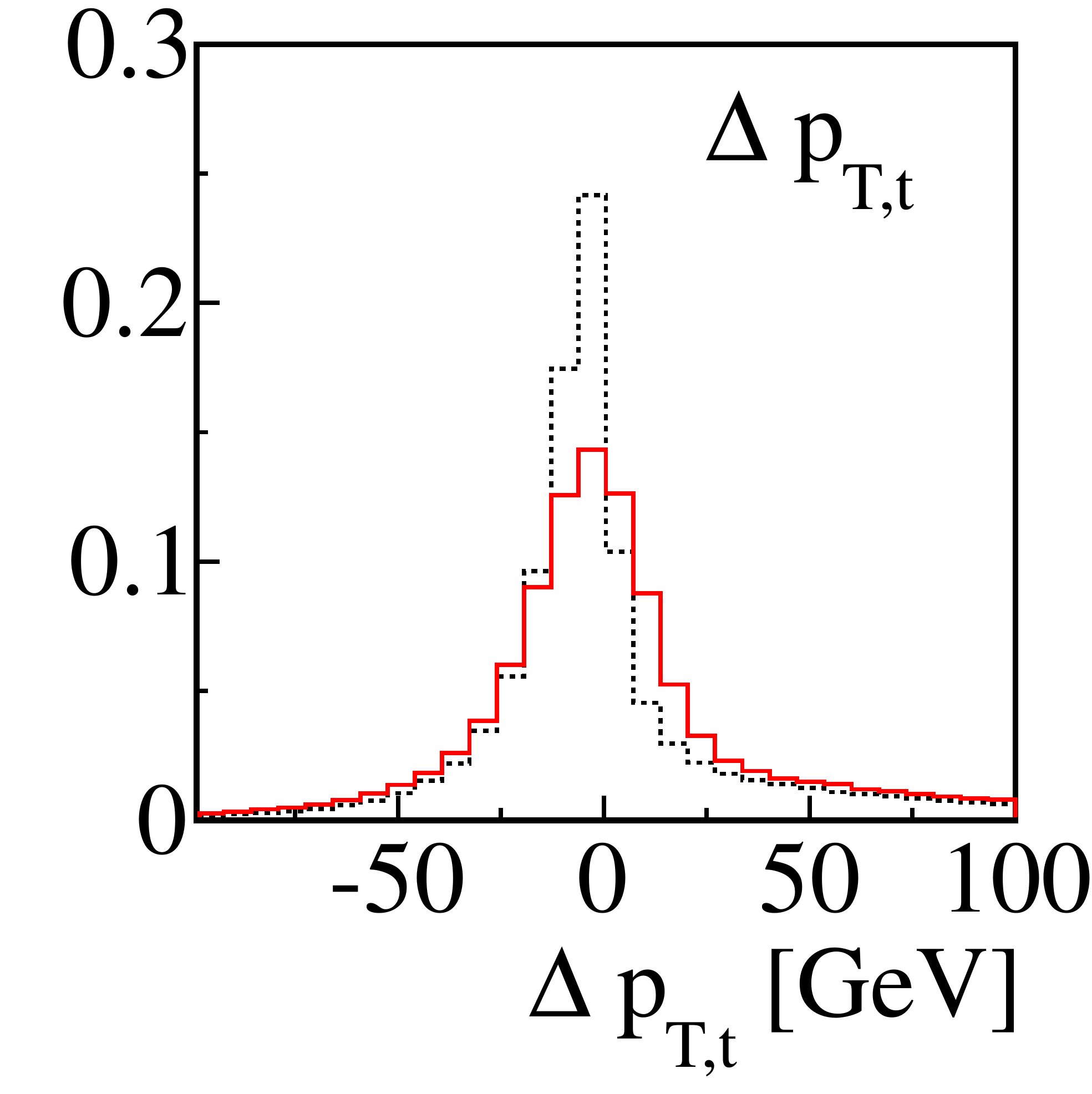}
\includegraphics[width=0.195\textwidth]{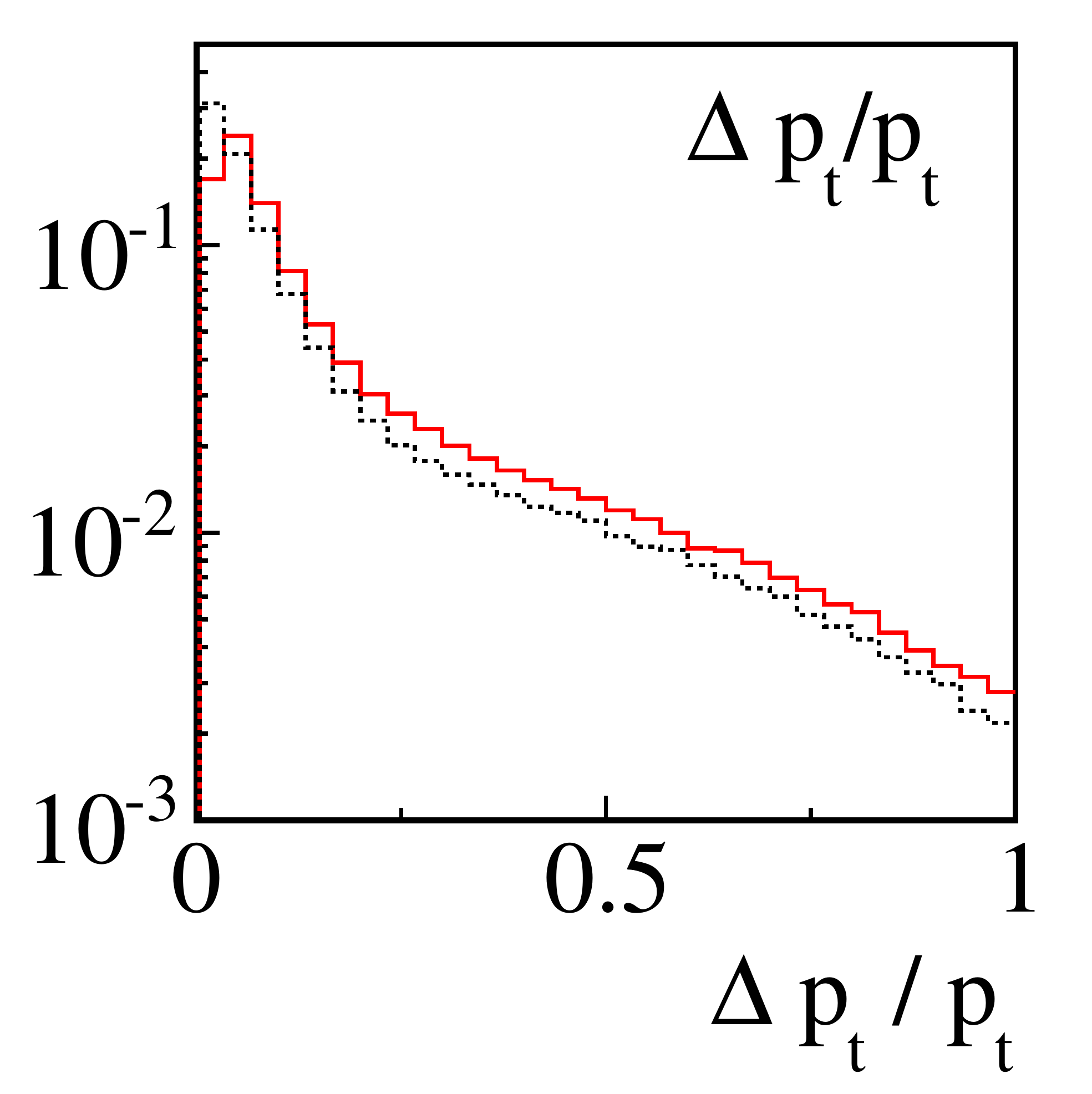}
\includegraphics[width=0.195\textwidth]{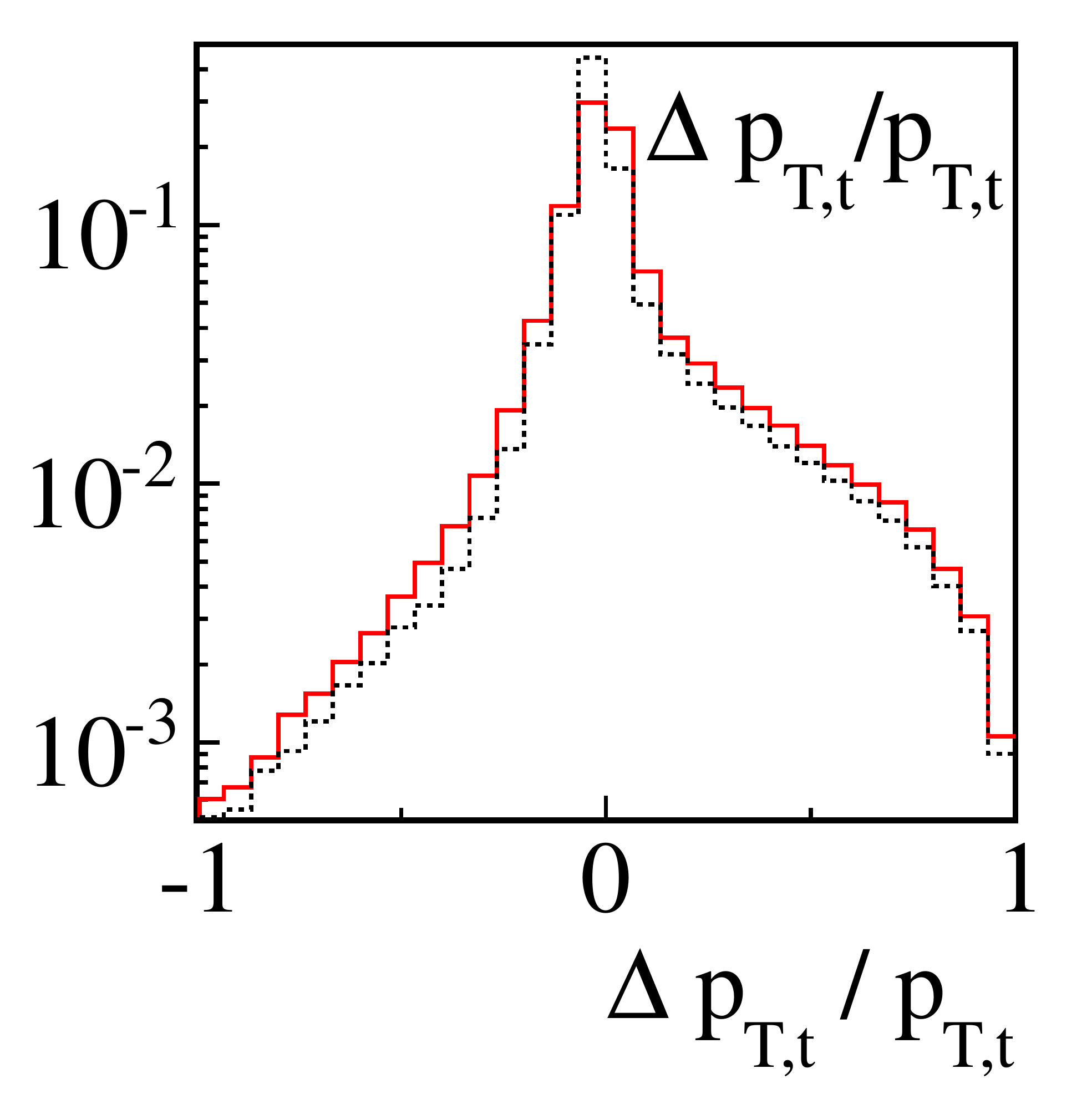}
\includegraphics[width=0.195\textwidth]{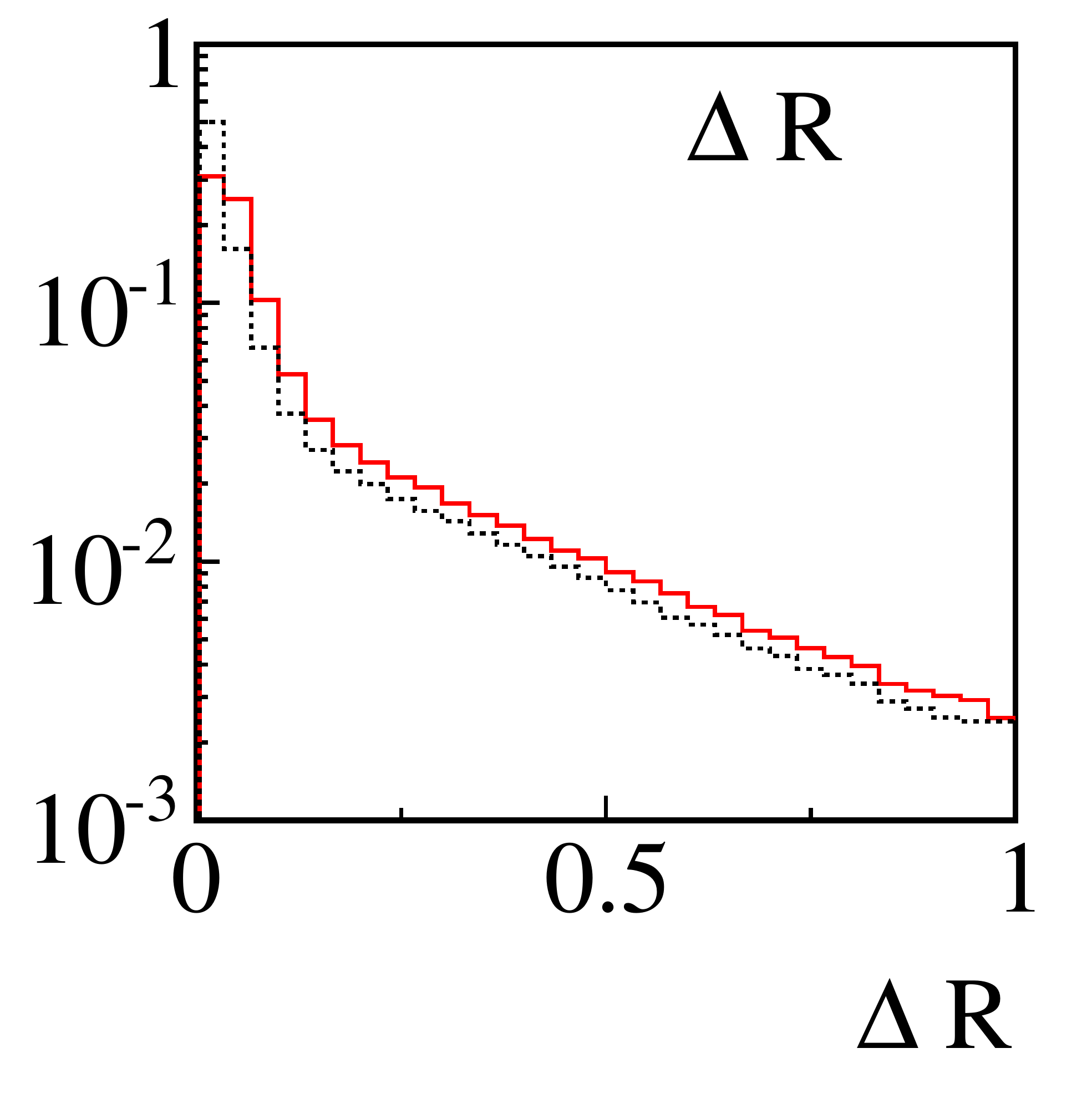}
\caption{From left to right, the $\Delta p_t$, $\Delta p_{T,t}$,
$\Delta p_t/|p_t^\text{tag}|$,  $\Delta p_{T,t}/p_{T,t}^\text{tag}$, and $\Delta R$
  distribution for the $t$-channel single top sample, shown with
  (solid red) and without {\sc Delphes} (dashed black).}
\label{fig:delphes}
\end{figure}

\section{Top-jet angles}

\begin{figure}[t]
\includegraphics[width=0.42\textwidth]{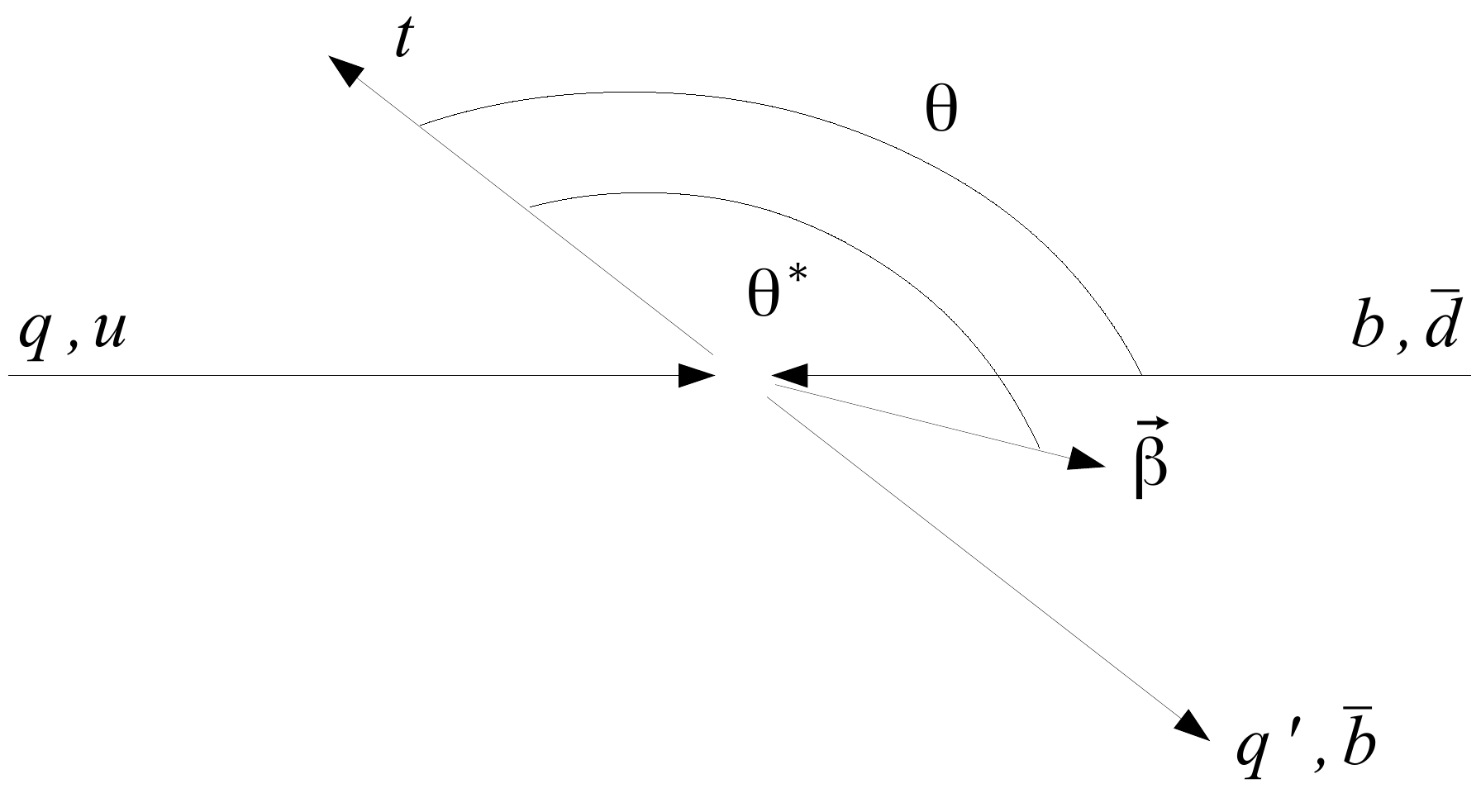}
\caption{Kinematics of single top production in the $s$- and
  $t$-channel with the definition of the angles $\cos\theta$ and
  $\cos\theta^*$.}
\label{fig:kinematics}
\end{figure}

The search for single tops using a top tagger has a significant
advantage: once the tagger has identified a top jet we automatically
get a full momentum reconstruction of this top quark. As described in
the previous appendix this 4-momentum reconstruction is quite
accurate. The obvious challenge is to define appropriate angular
variables which allow us to distinguish between $s$-channel and
$t$-channel single top production and the top pair background,
respectively.\medskip

For $t$-channel production the hard partonic process is $qb \to
q^\prime t$. Here, $q (q^\prime)$ denotes possible initial (final)
(anti-)quarks of the first and second generations. Assuming
(unrealistic) full control over the initial and final states we define
$\theta$ as the angle between the top direction $t$ and the incoming
quark direction $q$ in the center-of-mass frame, \ie the scattering
angle of the hard subprocess. The differential cross section with fixed center-of-mass energy squared $s$
gives its distribution
\begin{alignat}{5}
\frac{d\sigma_t}{d\cos\theta} \propto 
\dfrac{1}{\left( 1+\cos\theta +  \dfrac{2m_W^2}{s-m_t^2} \right)^2} \; ,
\end{alignat}
with a maximum at $\cos\theta=-1$ or $\theta = \pi$. This means
the top prefers to follow the direction of the incoming bottom, \ie
opposite to the incoming quark $q$. This is typical for
$t$-channel processes.  Note that for anti-top production $q\bar{b}
\to \bar{t}q^\prime $ we define $\theta$ as the angle between the
$\bar{t}$ and the incoming quark $q$ and obtain the same angular
distribution through charge conjugation.\medskip

\begin{figure}[t]
\includegraphics[width=0.24\textwidth]{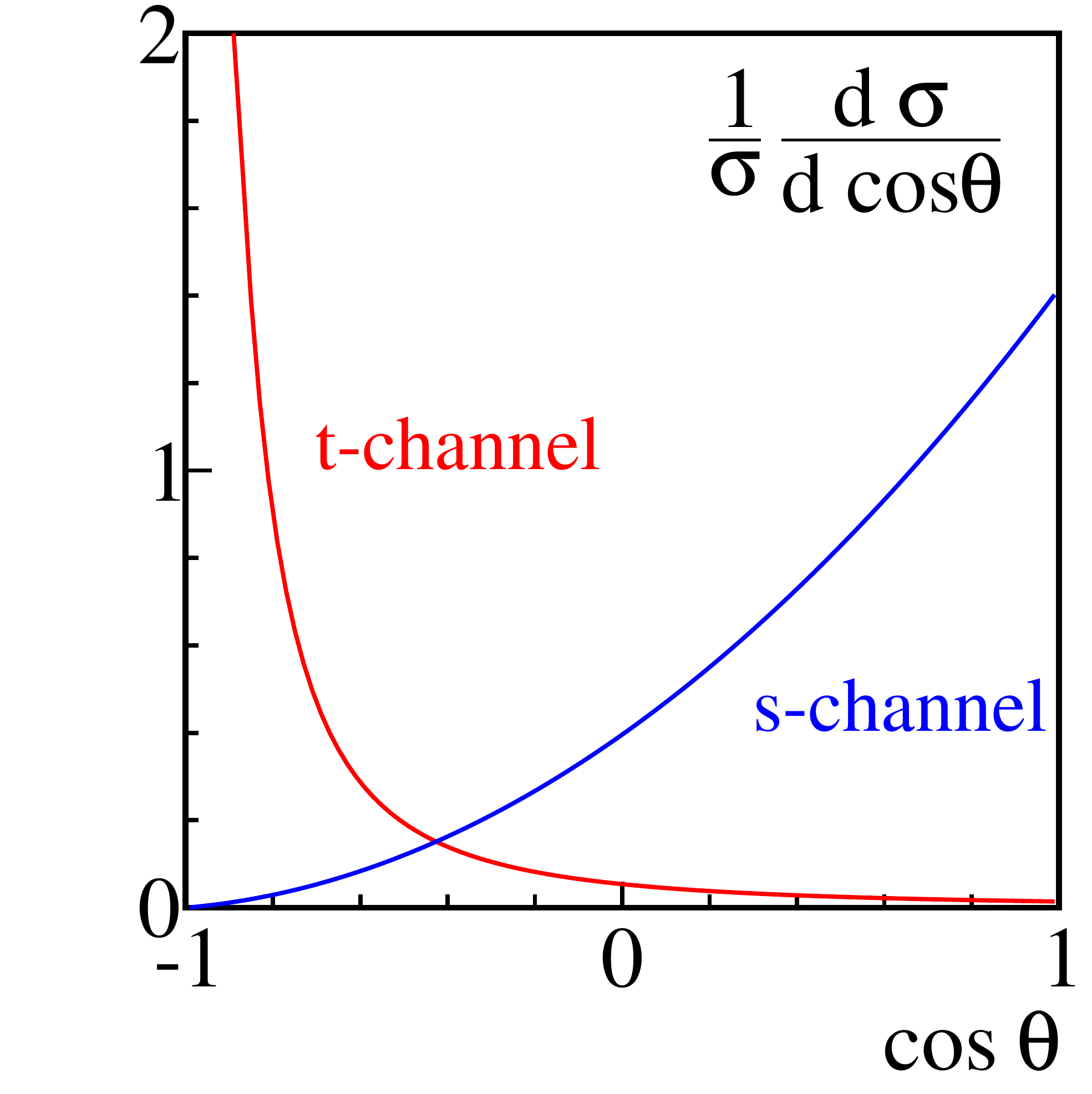}
\includegraphics[width=0.24\textwidth]{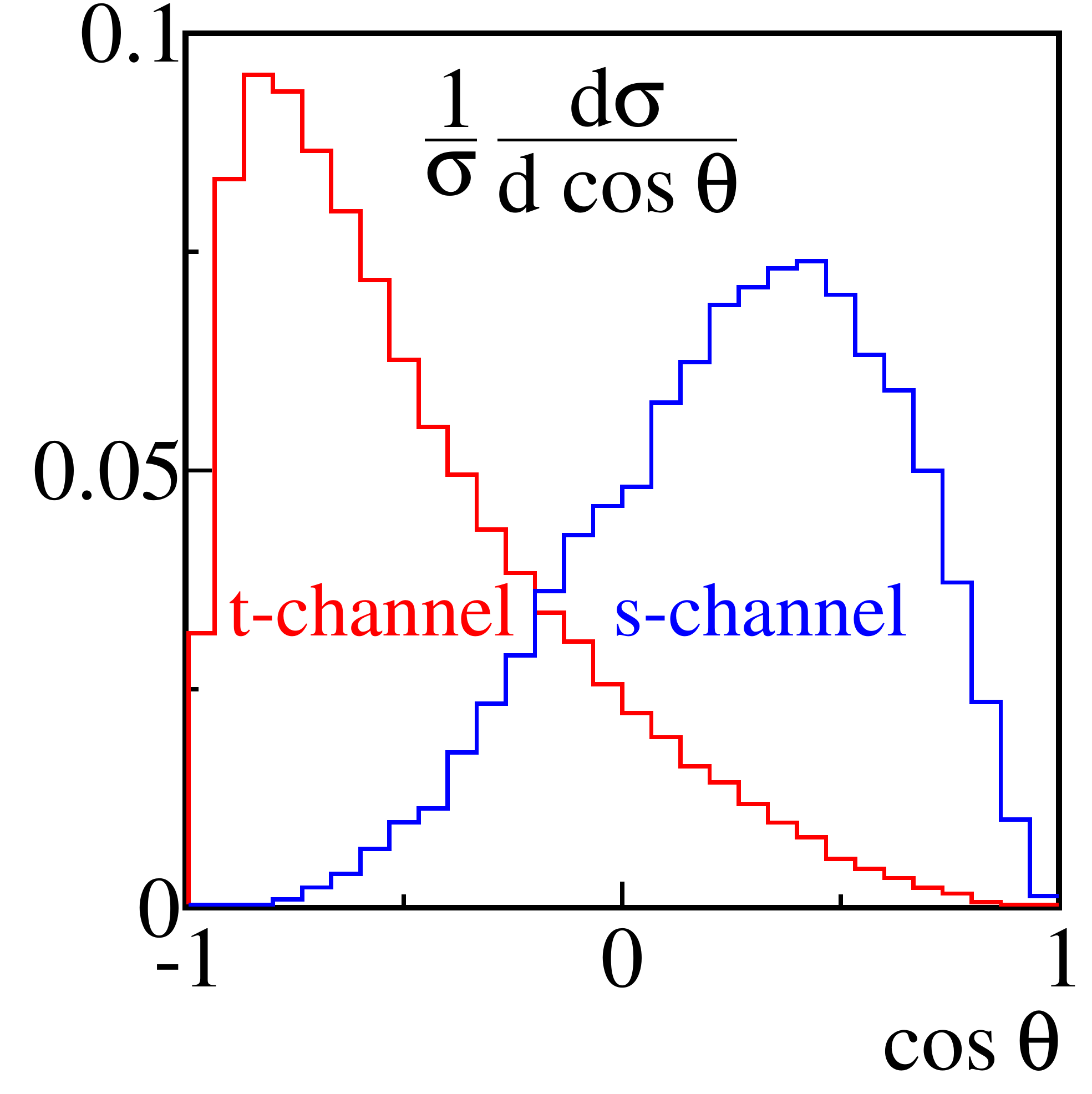}
\includegraphics[width=0.24\textwidth]{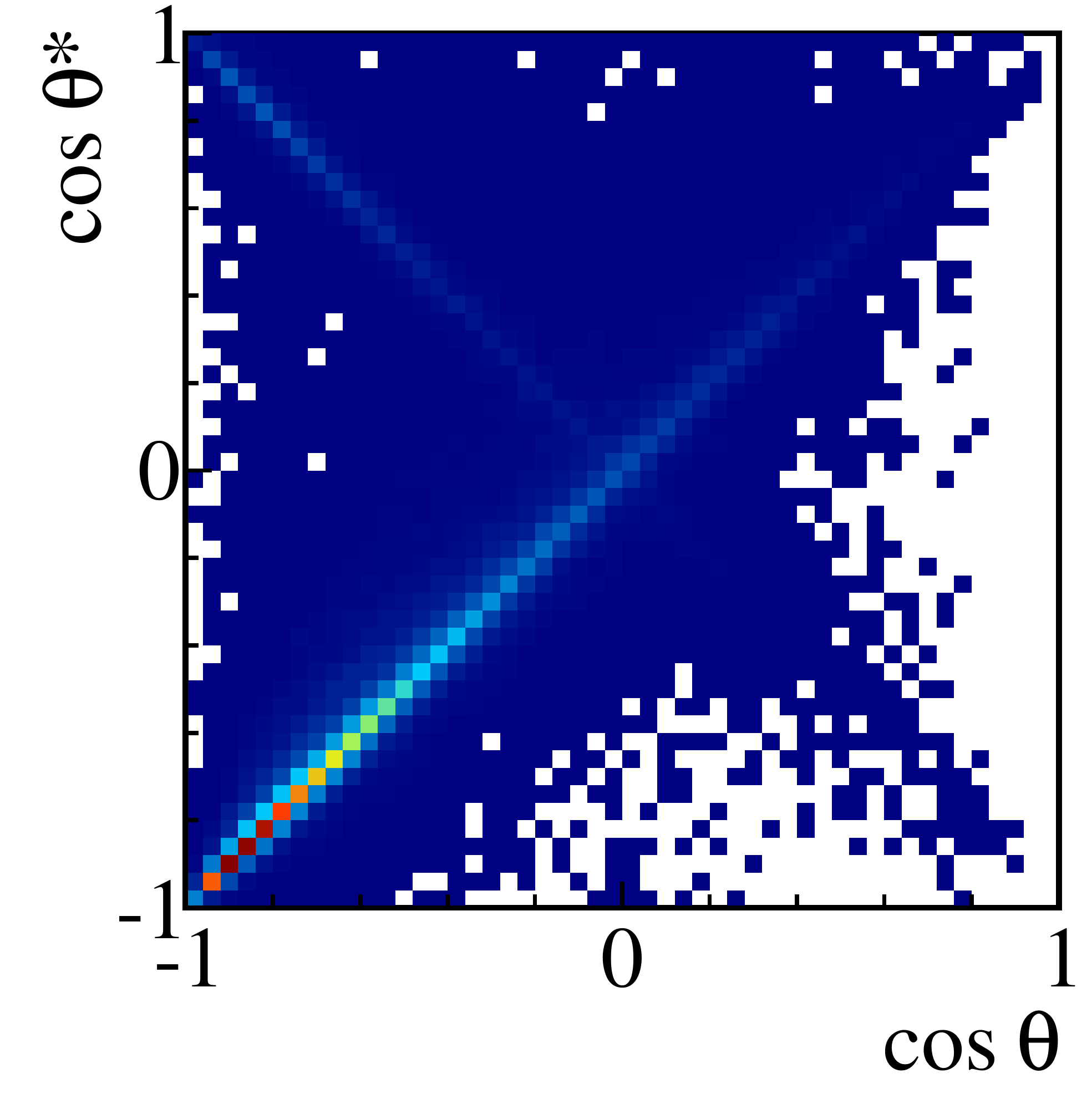}
\includegraphics[width=0.24\textwidth]{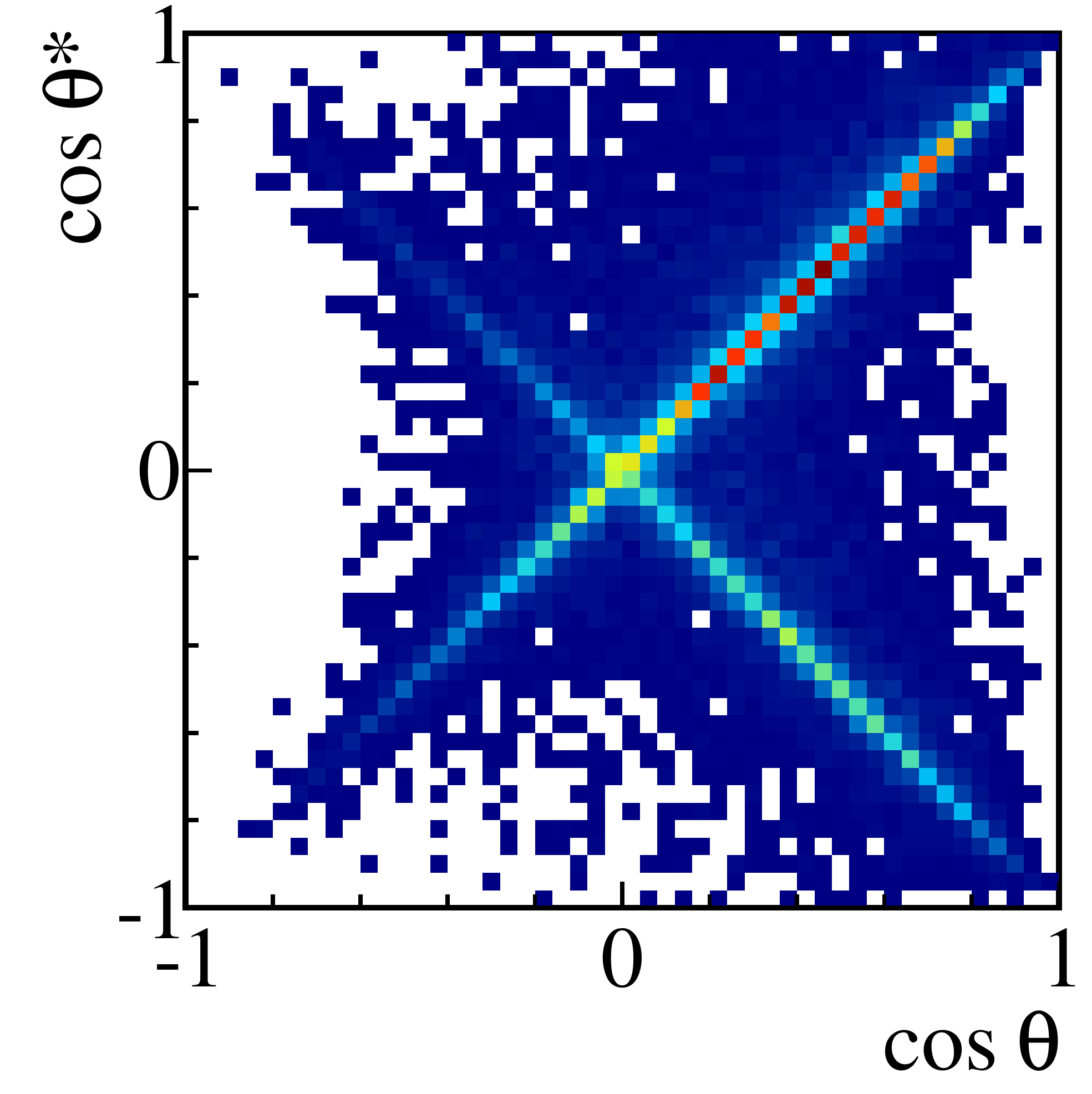}
\caption{From the left: $d\sigma_t/d\cos\theta$ and
  $d\sigma_s/d\cos\theta$ for a fixed partonic energy
  $\sqrt{s}=500$~GeV; $d\sigma_t/d\cos\theta$ and
  $d\sigma_s/d\cos\theta$ for single top events with $p_{T,t} >
  200$~GeV; $\cos\theta$ vs.  $\cos\theta^*$ for $t$-channel single
  top production; the same for $s$-channel single top production. The
  dense region is shown in red.}
\label{fig:costheta}
\end{figure}

For $s$-channel production $u\bar{d} \to t\bar{b}$ we define $\theta$
as the angle between the top direction and the incoming $u$ quark. The
differential cross section is
\begin{alignat}{5}
\frac{d\sigma_s}{d\cos\theta} \propto 
\dfrac{2m_t^2(1+\cos\theta) + s \left(1- \dfrac{m_t^2}{s} \right) (1+\cos\theta)^2}
      {s-m_W^2} \; .
\end{alignat}
The distribution has a maximum at $\cos\theta=1$ or $\theta=0$, \ie 
the top tends to be emitted in the incoming $u$-quark direction. 
For anti-top production $d\bar{u} \to \bar{t}b$ we
define $\theta$ as the angle between the anti-top and the incoming
$\bar{u}$ and obtain the same angular distribution 
through charge conjugation.\medskip

The left panel of Fig.~\ref{fig:costheta} shows the above
distributions as functions of $\cos\theta$ with fixed
$\sqrt{s}=500~\gev$.  In the central panel we see the same
distributions, now folded with parton densities and 
with an explicit $p_{T,t}$ cut, $p_{T,t}>200~\gev$. 
A loss of the events around $\cos\theta \sim \pm 1$
can be observed.  Note that cuts on $p_{T,t}$ and $\cos\theta$ are
linked because of the leading order kinematic relation
$p_T=(\sqrt{s}/2)|\cos\theta|$.\medskip

Unfortunately, the angle $\theta$ cannot be extracted at the LHC event
by event. Therefore, we define the modified angle $\theta^*$ between
the top direction in the top-jet rest frame and the direction of the
boost $\vec{\beta}$ from the rest frame to the laboratory frame,
as seen in Fig.~\ref{fig:kinematics}. This boost lies mostly in the direction of
the incoming beams and reflects the difference in the partonic
momentum fractions of the two incoming (anti-)quarks. The behavior of
$\theta^*$ closely tracks the above described angle $\theta$.

For $t$-channel single tops the boost vector is preferably pointed in
the initial quark direction, because incoming bottom partons have
significantly softer partonic energy spectra.  Therefore, the
$\cos\theta^*$ distribution essentially reproduces the $\cos\theta$
distribution as shown in Fig.~\ref{fig:cos}.  This relation for 
$t$-channel single top sample at the parton level is 
shown in the third panel of Fig.~\ref{fig:costheta}. 
We see a clear preference for
$\cos\theta^*=\pm \cos\theta$, where the more likely relative plus
sign appears when the incoming quark is more energetic than the
incoming bottom. Note that it is also true for 
$t$-channel anti-top production.
The same correlation appears for $s$-channel single
top sample as seen in the most right panel of Fig.~\ref{fig:costheta},
while more minus signs appear due to 
anti-top single production.


\end{document}